\documentclass[a4paper,fleqn,usenatbib]{mnras}

\usepackage{newtxtext,newtxmath}

\usepackage[T1]{fontenc}
\usepackage{ae,aecompl}

\usepackage{graphicx}	
\usepackage{amsmath}	
\usepackage{amssymb}	

\newcommand{\cmnt}[1]{}
\newcommand{\viva}[1]{#1}
\newcommand{\review}[1]{#1}

\newcommand{\program}[1]{\texttt{\sc #1}}

\newcommand{\porb}{P_\mathrm{orb}}
\newcommand{\halpha}{H$\alpha$}
\newcommand{\hbeta}{H$\beta$}
\newcommand{\hei}{\ion{He}{I}}
\newcommand{\heii}{\ion{He}{II}}

\newcommand{\filus}{\textit{u$_s$}}	
\newcommand{\filgs}{\textit{g$_s$}}	
\newcommand{\filrs}{\textit{r$_s$}}	
\newcommand{\filu}{\textit{u'}}	
\newcommand{\filg}{\textit{g'}}	
\newcommand{\filr}{\textit{r'}}	
\newcommand{\filkg}{\textit{KG5}}	

\newcommand{\warwick}{$^1$}
\newcommand{\sheffield}{$^4$}
\newcommand{\ioa}{$^3$}
\newcommand{\iac}{$^5$}
\newcommand{\eso}{$^6$}
\newcommand{\caldav}{$^2$}
\newcommand{\narit}{$^8$}
\newcommand{\lalaguna}{$^7$}

\title[Six Ultracompact Accreting Binaries]{Spectroscopic and Photometric Periods of Six Ultracompact Accreting Binaries}

\author[M. J. Green]{
Matthew J. Green\warwick\thanks{E-mail: mjgreenastro@gmail.com},
Thomas R. Marsh\warwick\thanks{E-mail: t.r.marsh@warwick.ac.uk},
Philip J. Carter\caldav,
Danny Steeghs\warwick,
\newauthor
Elm\'{e} Breedt\ioa,
V. S. Dhillon\sheffield$^,$\iac,
S. P. Littlefair\sheffield,
Steven G. Parsons\sheffield,
Paul Kerry\sheffield,
\newauthor
Nicola P. Gentile Fusillo\eso,
R. P. Ashley\warwick,
Madelon C. P. Bours\warwick,
Tim Cunningham\warwick,
\newauthor
Martin J. Dyer\sheffield,
Boris T. G\"{a}nsicke\warwick,
Paula Izquierdo\iac$^,$\lalaguna,
Anna F. Pala\eso,
\newauthor
Chuangwit Pattama\narit,
Sabrina Outmani\warwick,
David I. Sahman\sheffield,
Boonchoo Sukaum\narit,
\newauthor
and James Wild\sheffield.
\\
\warwick Astronomy and Astrophysics Group, Department of Physics, University of Warwick, Coventry, CV4 7AL, United Kingdom
\\
\caldav Department of Earth and Planetary Sciences, University of California Davis, One Shields Avenue, Davis, CA 95616
\\
\ioa Institute of Astronomy, University of Cambridge, Madingley Road, Cambridge, CB3~0HA, United Kingdom
\\
\sheffield Department of Physics and Astronomy, University of Sheffield, Sheffield, S3~7RH, United Kingdom
\\
\iac Instituto de Astrof\'isica de Canarias, 38205 La Laguna, Tenerife, Spain
\\
\eso European Southern Observatory, Karl Schwarzschild Stra{\ss}e 2, Garching, 85748, Germany
\\
\lalaguna Departamento de Astrof\'isica, Universidad de La Laguna, 38206 La Laguna, Tenerife, Spain
\\
\narit National Astronomical Research Institute of Thailand (Public Organization), 260 Moo 4, Donkaew, Mae Rim, Chiang Mai, 50180, Thailand
}

\date{Accepted XXX. Received YYY; in original form ZZZ}

\pubyear{2020}

\begin{document}
\label{firstpage}
\pagerange{\pageref{firstpage}--\pageref{lastpage}}
\maketitle

\begin{abstract}
Ultracompact accreting binary systems each consist of a stellar remnant accreting helium-enriched material from a compact donor star. 
Such binaries include two related sub-classes, AM\,CVn-type binaries and helium cataclysmic variables, in both of which the central star is a white dwarf.
We present a spectroscopic and photometric study of six accreting binaries with orbital periods in the range of 40--70\,min, including phase-resolved VLT spectroscopy and high-speed ULTRACAM photometry.
Four of these are AM\,CVn systems and two are helium cataclysmic variables.
For four of these binaries we are able to identify orbital periods (of which three are spectroscopic).
SDSS\,J1505+0659 has an orbital period of 67.8\,min, \review{significantly longer than previously believed}, and longer than any other known AM\,CVn binary.
We identify a WISE infrared excess in SDSS\,J1505+0659 that we believe to be the first direct detection of an AM\,CVn donor star in a non-direct impacting binary.
The mass ratio of SDSS\,J1505+0659 is consistent with a white dwarf donor.
CRTS\,J1028$-$0819 has an orbital period of 52.1\,min, the shortest period of any helium cataclysmic variable.
MOA\,2010-BLG-087 is co-aligned with a K-class star that dominates its spectrum.
ASASSN-14ei and ASASSN-14mv both show a remarkable number of echo outbursts following superoutbursts (13 and 10 echo outbursts respectively). 
ASASSN-14ei shows an increased outburst rate over the years following its superoutburst, perhaps resulting from an increased accretion rate.

\end{abstract}




\section{Introduction}

Cataclysmic variables (CVs) are compact, interacting binaries. Each CV consists of a white dwarf accreting material from a donor star which fills its Roche lobe \citep{Warner1995}.
For non-magnetic systems, this accretion occurs via an accretion disc around the central white dwarf.
In the vast majority of CVs, the donor closely resembles a main sequence star at the point at which accretion starts. For such systems there exists a minimum orbital period, \review{observed to be} $79.6 \pm 0.2$\,min \citep{Knigge2011,McAllister2019}.
Accreting binaries which have orbital periods shorter than this limit must therefore have donors which were not typical main-sequence stars at the point of first contact. 
The most common explanation is for the donor to be evolved, with a helium-rich nature and consequently greater density that allows the binary to exist at shorter orbital periods. 
CVs with evolved donors and periods shorter than the canonical period minimum can be divided into two sub-classes: helium-rich CVs (He CVs), which accrete helium-enriched material containing some amount of hydrogen, and AM Canum Venaticorum (AM\,CVn) type binaries, which accrete helium-dominated material with no detectable hydrogen.
AM\,CVn binaries include the shortest-period systems known \citep{Motch1996,Israel1999,Steeghs2006,Roelofs2010}, and are expected to be among the strongest Galactic sources of gravitational wave radiation in the LISA frequency range \citep{Kremer2017,Breivik2018,Kupfer2018}.
Although He~CVs and AM\,CVn binaries are similar in appearance, the extent to which they are evolutionarily related is not yet well understood.

The formation path of a He~CV is as follows.
In a volume-limited sample, 5$\pm$3 per cent of CVs appear to have evolved donors \citep{Pala2020}.
The majority of these donor stars still have hydrogen-dominated atmospheres and orbital periods longer than the canonical period minimum.
Over time, mass transfer will strip away the atmosphere of the donor star, revealing the helium-enriched core.
In this manner the accreted material becomes helium-rich. At the same time, the increasing density of the donor star will require that the binary evolves to a shorter orbital period.

The evolution of AM\,CVn binaries is uncertain, with three proposed channels (see \citealp{SolheimAMCVn} for an overview, and \citealp{Green2018b} and \citealp{Ramsay2018} for recent results).
In two of these channels, the binary goes through two phases of common envelope evolution \citep{Ivanova2013}, leaving either a double white dwarf binary \citep{Paczynski1967,Deloye2007} or a binary consisting of a white dwarf and a helium-burning star whose atmosphere has been stripped during the common envelope evolution \citep{Savonije1986,Iben1987,Yungelson2008}. 



In the third suggested channel of AM\,CVn evolution, the AM\,CVn binary descends directly from a He~CV in which the core of the donor star is sufficiently helium-enriched \citep{Podsiadlowski2003,Goliasch2015}.
As a source of AM\,CVn binaries, this channel has several uncertainties. 
Firstly, the region of initial parameter space for which a binary will be able to remove all detectable hydrogen from the accreted material is rather small, and many He~CVs will never successfully remove enough hydrogen to become AM\,CVn binaries \citep{Goliasch2015}. 
As such we might expect the number of known, sub-period minimum He~CVs to provide an approximate upper limit on the number of AM\,CVn binaries formed by this channel. 
Secondly, the majority of AM\,CVn binaries that do form by this channel are not expected to reach periods shorter than $\approx 45$\,min \citep{Podsiadlowski2003,Goliasch2015}. 
As a result, population synthesis models generally disfavour this model \textit{for short period AM\,CVn binaries} compared to the two previously discussed, predicting a contribution of order a few per cent to the population of AM\,CVn binaries with periods $\lesssim 30$\,min \citep{Nelemans2004,Goliasch2015}.
For AM\,CVn binaries with orbital periods $\gtrsim 45$\,min, however, this channel remains a viable option.
Indeed, the most precise to date mass and radius measurements of an AM\,CVn donor star (from the first fully-eclipsing AM\,CVn binary, Gaia14aae, which has an orbital period of 50\,min) agree best with predictions based on the He~CV channel \citep{Green2018,Green2019}.

Further attempts to study the prior evolution of these binaries are limited by the small numbers of such systems which are amenable to characterisation. 
Key properties to constrain the prior evolution of a binary are its orbital period ($\porb$) and the ratio of masses of its two stars ($q = M_2 / M_1$, where $M_1$ is the mass of the accretor and $M_2$ is the mass of the donor).
At present, 56 AM\,CVn binaries \citep{Ramsay2018} and 13 sub-period minimum He~CVs (see Table~\ref{tab:hecvs}) are known. Another two He~CVs are believed to be below the period minimum due to the low hydrogen content of their accreted material.
However, many systems of both classes are too faint for detailed follow-up.
Only 33 AM\,CVn binaries have published, directly measured orbital periods \citep[and references therein]{Green2018b}.
\review{Note that this number does not include those systems for which an indirect estimate of the orbital period has been made, based either on its superhump period (see the discussion in the following section) or on its outburst properties \citep{Levitan2015}.}
Only 17 AM\,CVn binaries \citep{Green2018b} and six sub-period minimum He~CVs (Table~1) have published mass ratio measurements.

In this paper, we present a spectroscopic and photometric study of six short-period accreting binaries, including four AM\,CVn binaries and two sub-period minimum He CVs (Table~\ref{tab:targets-6}). 
Our goal is to use these data to measure the orbital periods and, where possible, mass ratios of these systems, in the hope of exploring their prior evolution.

\begin{table}
\label{tab:hecvs}

\caption{Properties of known CVs (not including AM\,CVn binaries) with orbital periods shorter than 79.6\,min. 
The majority are He~CVs, but note that the donor of SDSS\,J1507+5230 does not appear to be evolved; instead, its orbital period can be explained by the low metallicity (and hence increased density) of its halo-population donor star. 
CRTS\,J0808+3550 and CRTS\,J1647+4338 do not have measured orbital periods, but are believed to be below the orbital period minimum due to the low hydrogen content of their accreted material.
In the case of V485\,Cen, $q$ is derived from spectroscopic studies of dynamics; for all other systems $q$ is derived from the superhump relationship. In these cases we have recalculated $q$ from the Stage B superhump excess using Equation~\ref{eq:mcallisterB}.
\textit{References:} [1] \citet{Kato2009}; [2] \citet{Woudt2012}; [3] \citet{Carter2013}; [4] \citet{Littlefield2013}; [5] \citet{Ramsay2014}; [6] \citet{Augusteijn1993}; [7] \citet{Augusteijn1996}; [8] \citet{Woudt2011}; [9] \citet{Thorstensen2002}; [10] \citet{Chochol2015}; [11] \citet{Imada2018}; [12] \citet{Breedt2012}; [13] \citet{Kato2015a}; [14] \citet{Szkody2005}; [15] \citet{Uthas2011}; [16] VSNET alert 19922; [17] VSNET alert 21021; [18] VSNET alert 21588; [19] \citet{Breedt2014}.
Note that a number of these measurements have not been confirmed by any peer-reviewed source.
}
\begin{tabular}{lccc}
Name & Period (min) & $q$ & Ref\\
\hline
CRTS\,J1028-0819	& 52.1		& $0.25 \pm 0.06$ 	& 1,2 \\
SBSS\,1108+574  	& 55.3		& $0.085 \pm 0.015$	& 3,4 \\
BOKS45906			& 56.4		& --		& 5 \\
V485\,Cen			& 59.0		& $0.38 \pm 0.13$  & 6,7 \\
CRTS\,J2333-1557	& 62.0		& -- 		& 8 \\
EI\,Psc				& 64.2		& $0.185 \pm 0.007$		& 9 \\
CRTS J1740+4147		& 64.8		& $0.055 \pm 0.006$		& 10,11 \\
CRTS\,J1122-1110	& 65.2		& $0.020 \pm 0.007$	& 12 \\
V418\,Ser			& 65.9      & -- 		& 13 \\
SDSS\,J1507+5230	& 67		& --		& 14,15 \\
ASASSN-16fy 		& 70.9		& --		& 16 \\
ASASSN-17gf			& 76.6		& -- 		& 17 \\
ASASSN-17ou			& 79.5		& -- 		& 18 \\
CRTS J0808+3550		& --		& --		& 19 \\
CRTS J1647+4338		& --		& --		& 19 \\
\hline
\end{tabular}

\end{table}

\subsection{Photometric Behaviour}

We start with a brief overview of the photometric behaviour of CVs and AM\,CVn binaries, as these binaries exhibit a range of phenomena on various timescales.

Both CVs and AM\,CVn binaries undergo brightening events known as dwarf nova outbursts. These outbursts are caused by an instability in the temperature and density of the accretion disc, causing the material to temporarily ionise, which in turn causes the system to brighten by $\sim$2-9\,mag.
This brightening can last of order days to weeks.
The outbursts of hydrogen- and helium-dominated discs are broadly similar, though different in some specifics \citep{Lasota2001,Kotko2012,Cannizzo2015}.
In short-period CVs and AM\,CVn binaries, the outbursts that are seen generally belong to the subtype of dwarf nova outbursts known as `superoutbursts' \citep{Kotko2012}. Superoutbursts generally last for weeks, and are brighter on average than a typical outburst.
Following a superoutburst, some CVs and AM\,CVn binaries show a series of `echo outbursts', which resemble ordinary dwarf nova outbursts. These are thought to be the result of reflections within the disc of the transition waves that mediate the change from ionised to neutral disc states \citep{Patterson1998a,Mayer2015}.

Among AM\,CVn binaries, there is a steep dependence between the orbital period of the binary and the mass transfer rate, which in turn drives the behaviour of the accretion disc.
Empirical relations exist between the outburst properties and the orbital period of the binary \citep{Levitan2015,Cannizzo2015,Cannizzo2019}. The tightest of these correlations, relating the recurrence time between outbursts to the orbital period, is given by
\begin{equation}
\tau_\mathrm{recur} = (1.53 \times 10^{-9}) P_\mathrm{orb}^{7.35} + 24.7 \mathrm{~days},
\end{equation}
based on a sample of 11 outbursting AM\,CVn binaries with orbital periods between 22 and 37\,min \citep{Levitan2015}.
Outbursts are rarely seen for systems with periods longer than 50\,min.

On timescales of minutes to hours, CVs and AM\,CVn binaries show a variety of photometric signals. Some of these may be quasi-periodic variability originating in the disc \citep{Fontaine2011,Kupfer2015}. 
However, many CVs and some AM\,CVn binaries show a photometric signature coincident with the orbital period of the binary \citep{Armstrong2012,Green2018b}. 
Alongside this, many AM\,CVn binaries and short-period CVs show a photometric signature known as the `superhump' period.
The signal is most often visible during superoutburst, and is coincident with a beat frequency between the orbital period and the precession period of an eccentric disc \citep{Patterson2005}. This superhump period is generally a few per cent longer than the orbital period.
The disc precession itself is sometimes seen photometrically \citep{Green2018b}, but as the timescale is of order 12\,hours such detections are often not possible from single-site, ground-based photometry.

The precession of the disc is driven by the tidal pull of the donor star. An empirical relationship exists between the superhump period, the orbital period and the mass ratio of the two stars \citep{Patterson2005,Knigge2006,Kato2013}. Various forms of the relationship are given in the literature. For this work, we will use the expression given by \citet{McAllister2019}. This expression includes uncertainties on the relationship itself, following the form of \citet{Knigge2006} but using a larger sample of systems. For stage B superhumps and a sample of 24 systems, \citet{McAllister2019} gives
\begin{equation}
q(\epsilon) = (0.118 \pm 0.003) + (4.45 \pm 0.28) \times (\epsilon - 0.025),
\label{eq:mcallisterB}
\end{equation}
where $\epsilon = (P_\mathrm{sh} - \porb) / \porb$, and $P_\mathrm{sh}$ and $\porb$ are the superhump and orbital periods, respectively. For stage C superhumps and a sample of 15 systems, they give a different relationship of
\begin{equation}
q(\epsilon) = (0.135 \pm 0.004) + (5.0 \pm 0.7) \times (\epsilon - 0.025).
\label{eq:mcallisterC}
\end{equation}
See \citet{Kato2009} for a discussion of the differences between superhump stages.
It should be noted that these relationships were calibrated based on hydrogen-accreting CVs, and doubts have been raised as to whether such a relationship can be extended to the helium-dominated discs of He~CVs or AM\,CVn binaries \citep{Pearson2007}. 
For the two AM\,CVn systems where $q$ has been measured by both superhump and alternate methods, one is in agreement and the other is not \citep{Roelofs2006,Copperwheat2011}.
Regardless, these relationships are widely used to characterise AM\,CVn binaries where other methods of measuring $q$ are not possible \citep{Roelofs2006,Armstrong2012,Isogai2019}.

\section{Observations}
\label{sec:observations-6}

\begin{table*}
\caption{Targets observed for this work. Magnitudes are taken from \textit{Gaia} DR2 \citep{GaiaCollaboration2018} and are given in the \textit{G} band, except where otherwise stated. Parallaxes are from \textit{Gaia} DR2, and distances are calculated using the method described by \citet{Kupfer2018}.
We have avoided the \textit{Gaia} magnitude of ASASSN-14ei as it is significantly discrepent from the CRTS \textit{V} magnitude (at \textit{G}=16.4).
See Section~\ref{sec:14mv-ll} for a discussion of this discrepancy.}
\label{tab:targets-6}
\begin{tabular}{lcccccc}
Target & Coordinates \review{(J2000)} & Class & \review{Quiescent} Mag. & Outburst & Parallax (mas) & \review{Distance (pc)}\\
\hline
SDSS\,J1505+0659		& 15:05:51.58 +06:59:48.7	& AM\,CVn & 19.1  & -- & 6.3(5) & $160 \pm 12$\\ 

ASASSN-14ei \review{= OX Eri}				& 02:55:33.39 $-$47:50:42.0	& AM\,CVn & \textit{V}=18.0 & \textit{V}=11.9 & 3.92(5) & $255 \pm 4$\\ 

ASASSN-14mv \review{= V493 Gem}				& 07:13:27.28 +20:55:53.4 & AM\,CVn & 18.0 & \textit{V}=12.8 & 4.07(12) & $247 \pm 7$\\
MOA\,2010-BLG-087		& 18:08:34.85 $-$26:29:22.8		& AM\,CVn & 19.4 & \textit{V}$\approx$14.7 & -- & --\\ 


CRTS\,J1028$-$0819		& 10:28:42.89 $-$08:19:26.6	& He CV & 19.4 & \textit{V}=14.8 & 1.4(3) & $720 \pm 200$\\ 

V418\,Ser				& 15:14:53.64 +02:09:34.5		& He CV & 20.2 & \textit{V}=15.8 & 0.5(6) & $1200 \pm 300$\\ 
\hline

\end{tabular}
\end{table*}

\begin{table*}
\caption{Observations presented in this paper. For spectroscopic observations we give the wavelength range, while we give the filter for photometric observations.}
\label{tab:6-observations}
\begin{tabular}{lcccc}
Target & Instrument & Date & \review{Wavelength range or} filter & Exposure (s)\\
\hline
\textit{Spectroscopy}\\

SDSS\,J1505+0659		& VLT + X-shooter	& 2015 Apr 9 		& 3200--10100\,\AA 				& 16 $\times$ 300 \\ 

ASASSN-14ei				& VLT + X-shooter	& 2015 Aug 11 	& 3200--10100\,\AA 				& 22 $\times$ 240 \\ 

ASASSN-14mv				& WHT + ISIS		& 2019 Feb 23		& 3850--4600\,\AA, 5750--9000\,\AA 				& 2 $\times$ 800 + 5 $\times$ 600 \\ 

MOA\,2010-BLG-087		& VLT + FORS2		& 2015 May 14 		& 3850--6300\,\AA 				& 10 $\times$ 240 \\ 


CRTS\,J1028$-$0819		& VLT + X-shooter	& 2015 Apr 9 		& 3200--10100\,\AA 				& 18 $\times$ 300 \\ 

V418\,Ser				& VLT + FORS2		& 2015 Apr 9 		& 3850--6300\,\AA 				& 21 $\times$ 240 \\ 

\\

\textit{Photometry}\\

SDSS\,J1505+0659		& TNT + ULTRASPEC	& 2014 Jan 27		& \filkg				& 207 $\times$ 10	\\
						& NTT + ULTRACAM	& 2019 Mar 28		& \filus\filgs\filrs 	& 616 $\times$ 10 	\\

ASASSN-14ei				& NTT + ULTRACAM	& 2018 Jan 18 		& \filus\filgs\filrs 	& 785 $\times$ 10 	\\
						&					& 2018 Nov 09		& \filus\filgs\filrs 	& 2782 $\times$ 5 	\\
						&					& 2018 Nov 10		& \filus\filgs\filrs 	& 3030 $\times$ 5 	\\
						&					& 2018 Nov 12		& \filus\filgs\filrs 	& 2952 $\times$ 5 	\\
						
ASASSN-14mv				& TNT + ULTRASPEC	& 2015 Jan 04 		& \filg					& 6910 $\times$ 1	\\
						&					& 2015 Jan 05 		& \filg					& 1923 $\times$ 1 + 5047 $\times$ 0.5	\\
						&					& 2017 Dec 14 		& \filkg				& 754 $\times$ 10	\\
						&					& 2018 Feb 08		& \filkg				& 1346 $\times$ 10	\\
						&					& 2018 Feb 09 		& \filkg				& 1278 $\times$ 10	\\

MOA\,2010-BLG-087		& NTT + ULTRACAM	& 2010 Apr 21		& \filu\filg\filr		& 502 $\times$ 5.6	\\
						&					& 2010 Apr 22		& \filu\filg\filr		& 1866 $\times$ 5.6	\\
						&					& 2010 Apr 24		& \filu\filg\filr		& 1932 $\times$ 5.6	\\

V418\,Ser				& NTT + ULTRACAM	& 2018 Apr 15		& \filus\filgs\filrs 	& 597 $\times$ 15 	\\

\hline
\end{tabular}
\end{table*}

The targets for these observations were selected from known but poorly characterised accreting binaries believed to have orbital periods shorter than the period minimum. While all have been included in previously published lists of short-period binaries \citep[eg.][]{Breedt2015,Ramsay2018}, several have no peer-reviewed data available and none have been the subject of detailed follow-up before now.
A summary of all observations presented in this work is presented in Table~\ref{tab:6-observations}.

\subsection{VLT Spectroscopy}

Five targets were observed using the 8.2\,m Very Large Telescope (VLT) at Paranal Observatory, Chile, under observing proposal 095.D-0888.
Fainter targets ($G > 19.5$) were observed by FORS2 (the FOcal Reducer and low dispersion Spectrograph), a spectrograph with lower dispersion \citep{Appenzeller1998}.
Brighter targets were observed by X-shooter, an Echelle spectrograph \citep{Vernet2011}.
Each target was observed continuously for two hours.

Observations with FORS2 were carried out using the 600B grism and a slit width of 1", giving a resolution ($\lambda / \Delta \lambda$) of 780. Spectra were reduced using \texttt{\sc pamela} and \texttt{\sc molly} \citep{molly}\footnote{http://deneb.astro.warwick.ac.uk/phsaap/software/}.
Each frame was bias-subtracted and flat-corrected using tungsten lamp flats.
Sky lines were subtracted, and fluxes were extracted using optimal weights \citep{Marsh1988}.
Wavelength calibrations were performed using a HgCd arc lamp observed the preceding day.
Data were flux calibrated against spectrophotometric standard stars observed the same night: LTT\,7379 in both cases.

X-shooter observations were carried out using a slit width of 1.2" in the VIS arm (5500 to 10200\,\AA) and 1" in the UVB arm (3200 to 5500\,\AA), producing resolutions ($\lambda / \Delta \lambda$) of 6500 and 5400 respectively. Data from the NIR arm of X-shooter were not used due to the blue nature of these objects.
X-shooter data were reduced using the standard X-shooter pipeline produced by ESO (\texttt{\sc reflex}).

\subsection{WHT Spectroscopy}

A further target, ASASSN-14mv, was observed using the Intermediate-dispersion Spectrograph and Imaging System (ISIS) instrument on the 4.2\,m William Herschel Telescope (WHT).
This target was observed for approximately 75 minutes. 
ISIS has two optical arms with adjustable central wavelengths. 
One spectrum was taken with the arms centred on 3900\,\AA\ and 8200\,\AA, to search for calcium and nitrogen, and a further six with the arms centred on 4500\,\AA\ and 6600\,\AA.
The R600B and R600R gratings were used, with a GG495 order-sorting filter in the red arm. 
The slit width was 1.2", giving a resolution of 9200.
These spectra were also reduced using \texttt{\sc pamela} and \texttt{\sc molly}, using the same procedure as described above for FORS2. 
Wavelength calibrations were carried out using a CuNe+CuAr arc lamp observed before and after the target, and the flux was calibrated using the standard star LTT\,7379.

\subsection{Survey Photometry}

To characterise the outburst behaviour of these systems, we retrieved data from public-access photometric surveys, including
the American Association of Variable Star Observers (AAVSO),
the Catalina Real-time Transient Survey \citep[CRTS,][]{Drake2009},
the Palomar Transient Factory \citep[PTF,][]{Rau2009},
the Zwicky Transient Facility \citep[ZTF,][]{Masci2019},
the Panoramic Survey Telescope and Rapid Response System \citep[Pan-STARRS,][]{Chambers2016},
and the All-Sky Automated Search for SuperNovae \citep[ASAS-SN,][]{Shappee2014,Kochanek2017}.

Observations of several of our targets have previously been reported through the Variable Star Network \citep[VSNET,][]{Kato2004}. We reference the appropriate VSNET alerts where relevant.

\subsection{High-Speed Photometry}

Where possible, we collected high-speed photometry of these targets using the cameras ULTRACAM and ULTRASPEC \citep{ULTRACAM,ULTRASPEC}.
ULTRACAM is a triple-beam imaging photometer which uses frame-transfer CCDs to reduce dead-time between exposures to negligible amounts ($\approx 25$\,ms).
For these observations, it was mounted on the 3.5\,m New Technology Telescope (NTT) at La Silla Observatory, Chile.
For earlier observations the standard Sloan \filu\filg\filr\ filters were used in the three beams. For more recent observations, custom \filus\filgs\filrs\ filters were used \citep{Dhillon2018}, designed to replicate the wavelength ranges of the standard filters but with a higher throughput (especially in the \filus\ band).
ULTRASPEC is a single-band imaging photometer which also uses frame-transfer CCDs. It is mounted on the 2.5\,m Thai National Telescope (TNT) on Doi~Inthanon, Thailand.
For some observations the \filg\ band was used, while for other observations we used the custom \filkg\ filter, a broad filter approximately equal to \filu+\filg+\filr, chosen for its greater throughput \citep[see][Appendix A]{Hardy2017}.

Data were reduced using the standard \program{ultracam} pipeline. Each frame was bias- and dark-subtracted. Pixel sensitivity was corrected for using sky flats observed on the same night where available, otherwise using sky flats observed at another time during the observing run. 
Fluxes were extracted using a variable-width aperture of radius 1.7~$\times$ the full-width half-maximum of the point spread function of stars in that frame, with pixel fluxes weighted optimally \citep{Naylor1998}.
Each flux was divided by that of a nearby, non-variable comparison star to correct for atmospheric transparency variations. 
Solar System barycentric time corrections were applied to account for light travel time and time dilation.

The fluxes of the comparison stars were taken from the Sloan Digital Sky Survey (SDSS) or from the AAVSO Photometric All-Sky Survey (APASS).
Comparison stars used were at 15:05:49.3 +06:57:02.7 ($g' = 15.90$\,mag from SDSS) for SDSS\,J1505+0659; 02:55:33.1 $-$47:53:10.9 ($g' = 15.94$\,mag from APASS) for ASASSN-14ei; 07:13:22.8 +20:55:17.5 ($g' = 16.22$\,mag from APASS) for ASASSN-14mv; and 15:15:00.6 +02:09:05.6 ($g' = 16.35$\,mag from SDSS) for V418\,Ser.
For MOA\,2010-BLG-087, we used a comparison star at 18:08:35.3 $-$26:29:14.7 for atmospheric corrections. As no SDSS or APASS magnitude is available due to the crowded nature of the field, we flux-calibrated these data using a standard star (Ru\,152) observed on the same night.
Note that for the \filus\filgs\filrs\ filters used for much of the data, flux-calibration is only approximate: this is especially true for the \filus\ band, where magnitudes can differ by $\sim 0.1$\,mag from the corresponding \filu\ magnitude. As our photometric data are used only for timing purposes, such approximate calibrations are sufficient.

\begin{figure*}
\begin{center}
\includegraphics[width=500pt]{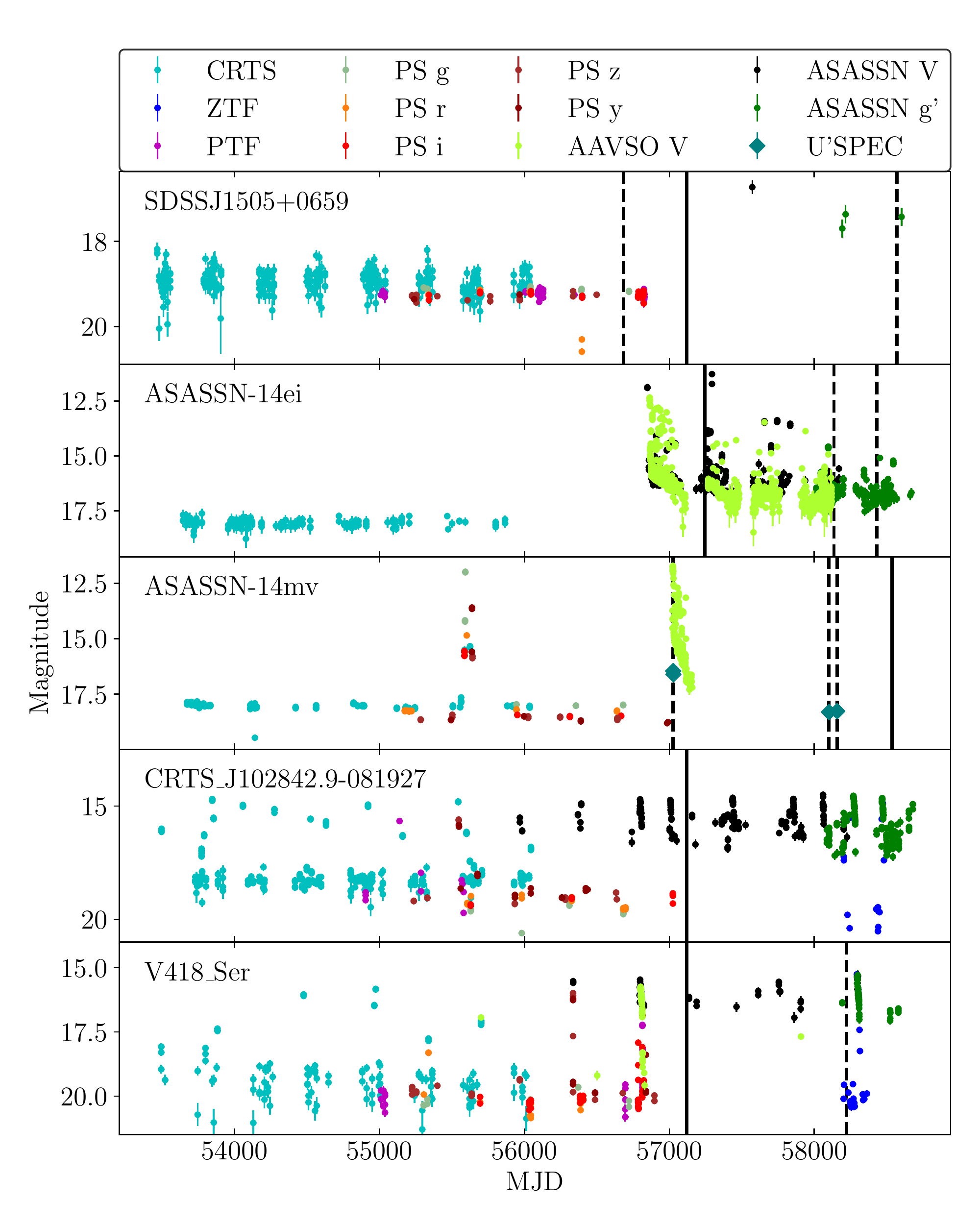}
\caption{Long-term lightcurves of the systems presented in this paper, except for MOA 2010-BLG-087. Vertical lines represent the times of our observations, either spectroscopy (solid) or high-speed photometry (dashed). Note that we have not shown the the ASAS-SN detections of ASASSN-14mv, due to the spurious 1.5\,mag offset discussed in Section~\ref{sec:14mv-ll}. Several targets have spurious ASAS-SN detections that coincide with the limiting magnitude of $\sim 17$; \review{most notably this includes the bright points of SDSS\,J1505+0659}.
}
\label{fig:longterm-lc}
\end{center}
\end{figure*}

\begin{figure*}
\begin{center}
\includegraphics[width=450pt]{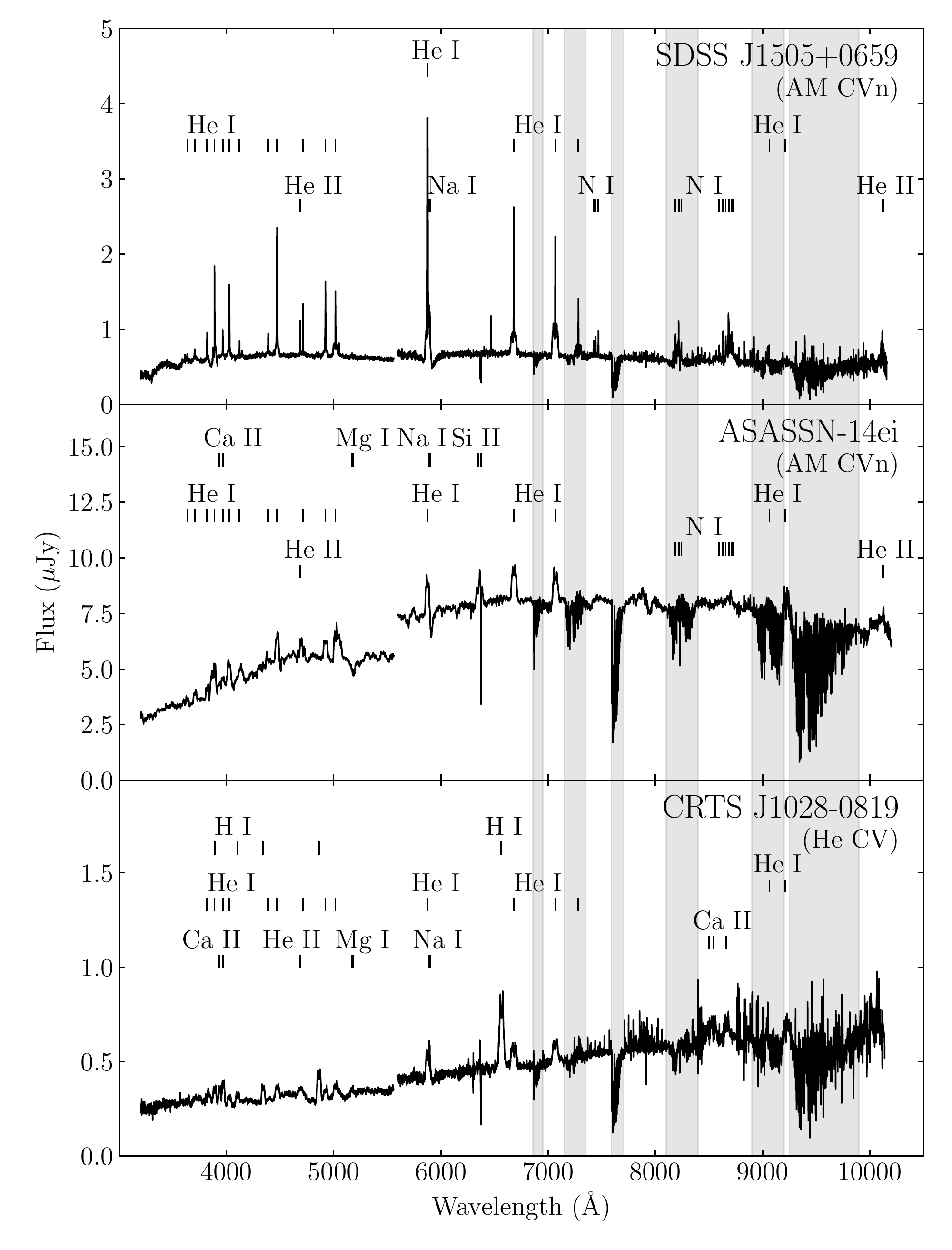}
\caption{Average spectra of systems observed with X-Shooter on the 8.2\,m VLT. Identified spectral features have been labelled. 
Grey shaded regions are dominated by telluric absorption, which has not been corrected for. Data have been rebinned by a factor of 5.
}
\label{fig:avspec-xshooter}
\end{center}
\end{figure*}

\begin{figure*}
\begin{center}
\includegraphics[width=450pt]{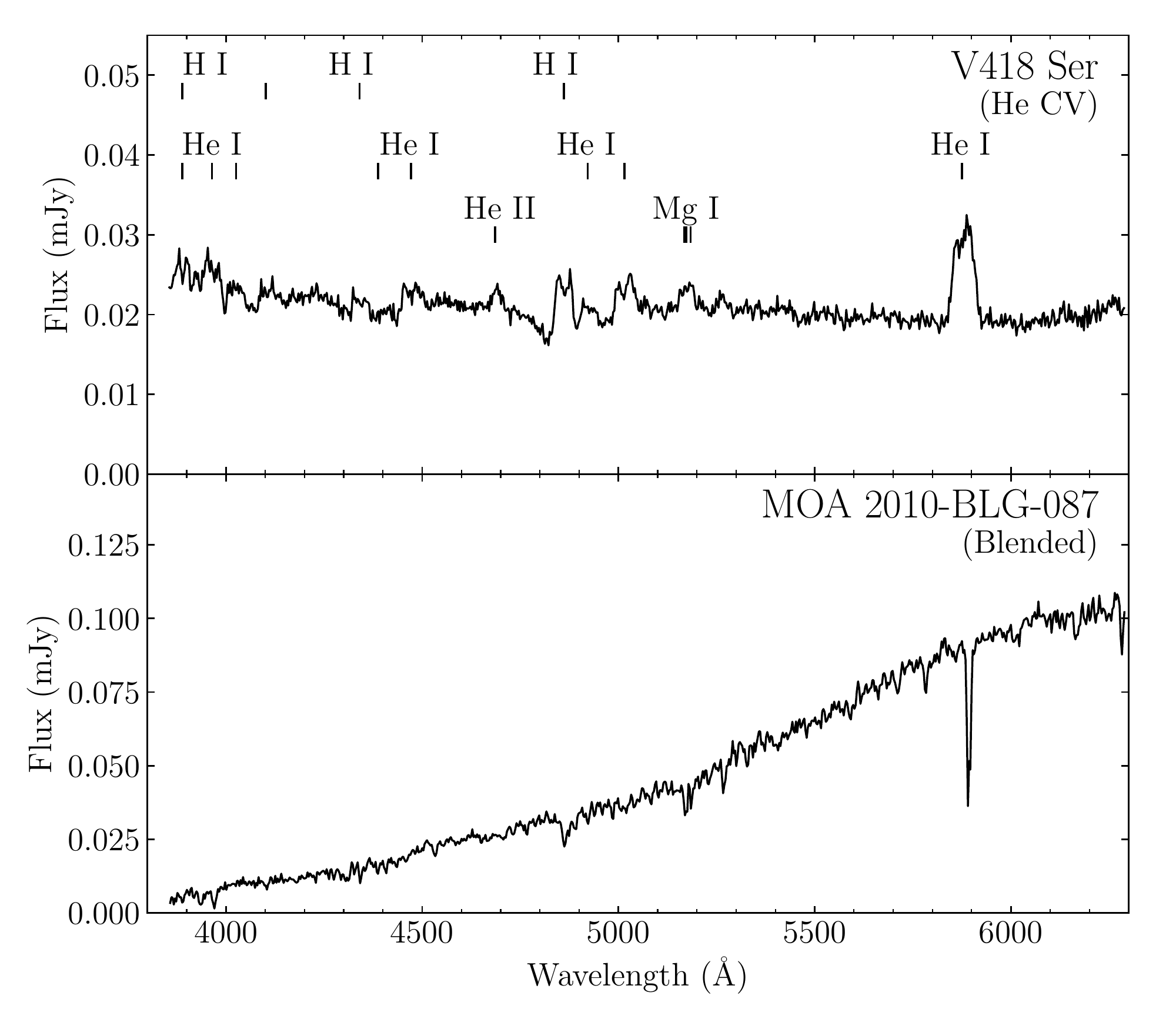}
\caption{Average spectra of systems observed with FORS2 on the 8.2\,m VLT. Identified spectral features have been labelled. 
As discussed in the text, MOA\,2010-BLG-087 is blended with a nearby G- or K-star that dominates the spectrum. Data have been rebinned by a factor of 2.
}
\label{fig:avspec-fors2}
\end{center}
\end{figure*}

\begin{figure*}
\begin{center}
\includegraphics[width=450pt]{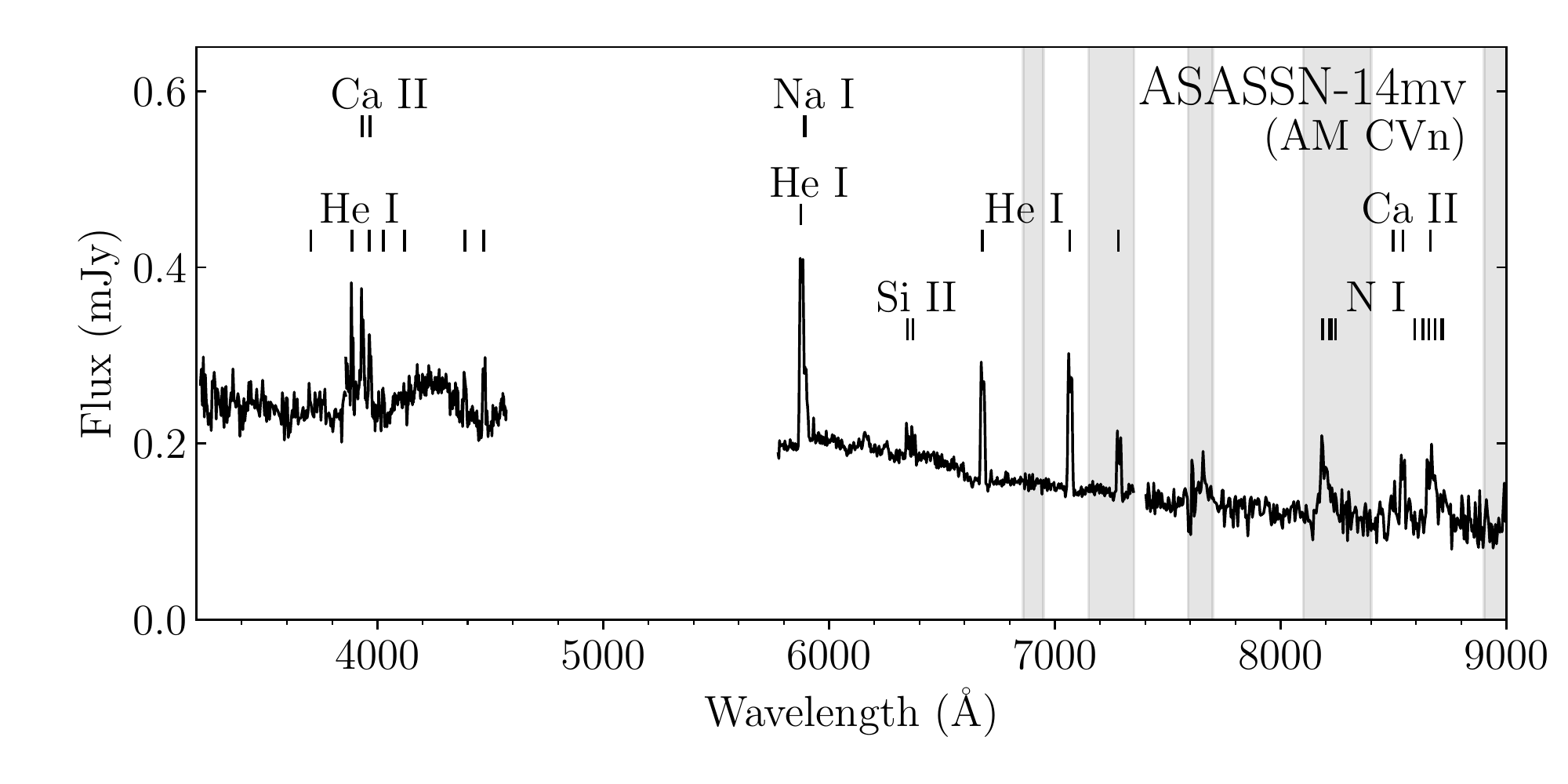}
\caption{Average spectrum of ASASSN-14mv, observed with ISIS on the 4.5\,m WHT. Identified spectral features have been labelled. 
Grey shaded regions are dominated by telluric absorption, which has been corrected for. Data have been rebinned by a factor of 5.
}
\label{fig:avspec-14mv}
\end{center}
\end{figure*}

\begin{figure}
\begin{center}
\includegraphics[width=\columnwidth]{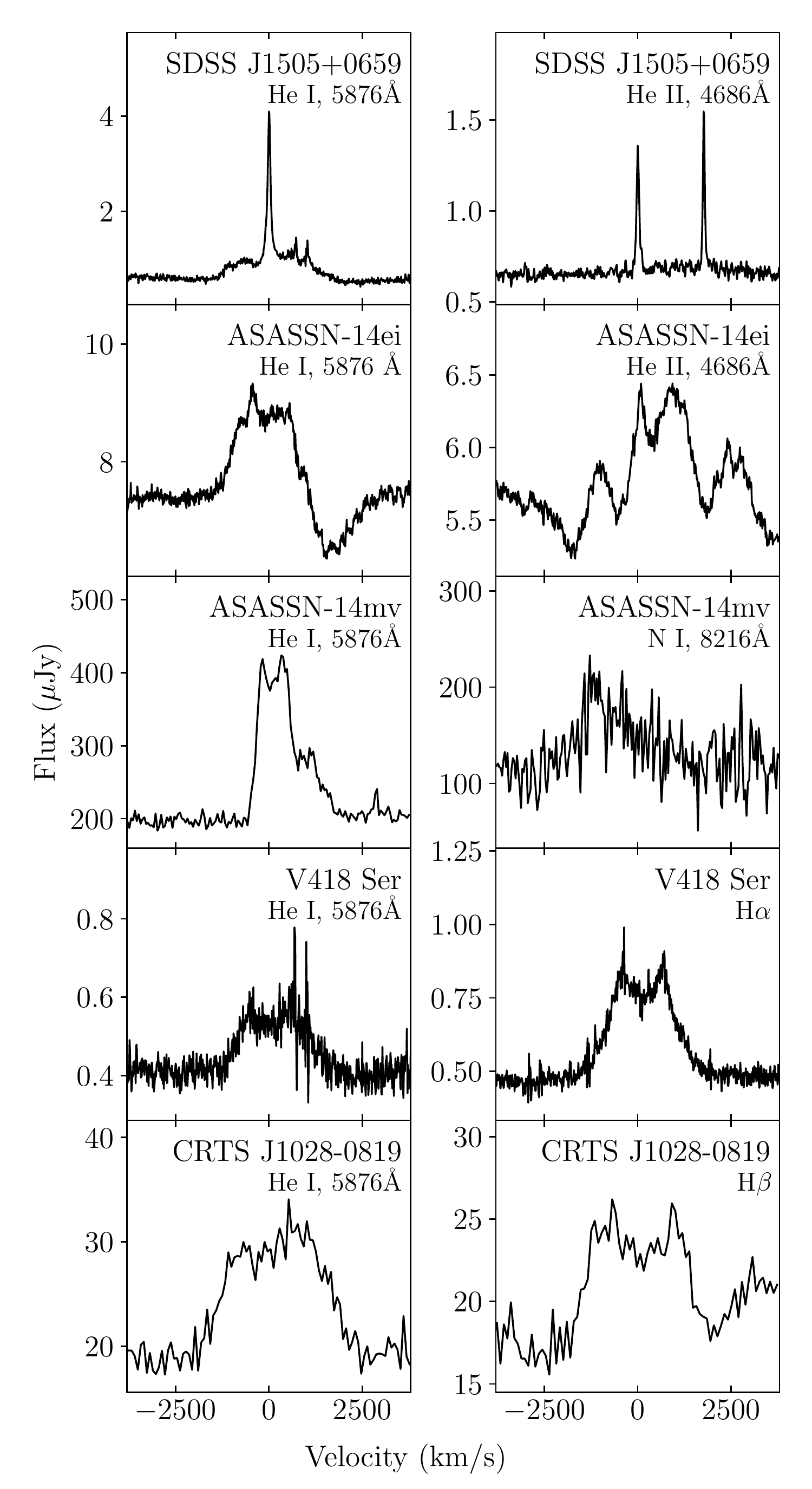}
\caption{Profiles of \hei\ 5876\,\AA\ lines for all five systems (left column), and of other spectral features of interest (right column).
In windows centred on the \heii\ 4686\,\AA\ line, the \hei\ 4713\,\AA\ line can also be seen.
}
\label{fig:profiles-6}
\end{center}
\end{figure}

\section{Results}
\label{sec:6-results}

Long-term lightcurves of each system are shown in Figure~\ref{fig:longterm-lc}.
Average spectra of each object are shown in Figures~\ref{fig:avspec-xshooter} (objects observed with X-shooter), \ref{fig:avspec-fors2} (objects observed with FORS2), and \ref{fig:avspec-14mv} (the object observed with ISIS). 
In Figure~\ref{fig:profiles-6}, we show line profiles for the 5876\,\AA\ \ion{He}{I} line in each system, along with other lines of interest.


\subsection{SDSS\,J1505+0659}

\begin{figure}
\begin{center}
\includegraphics[width=\columnwidth]{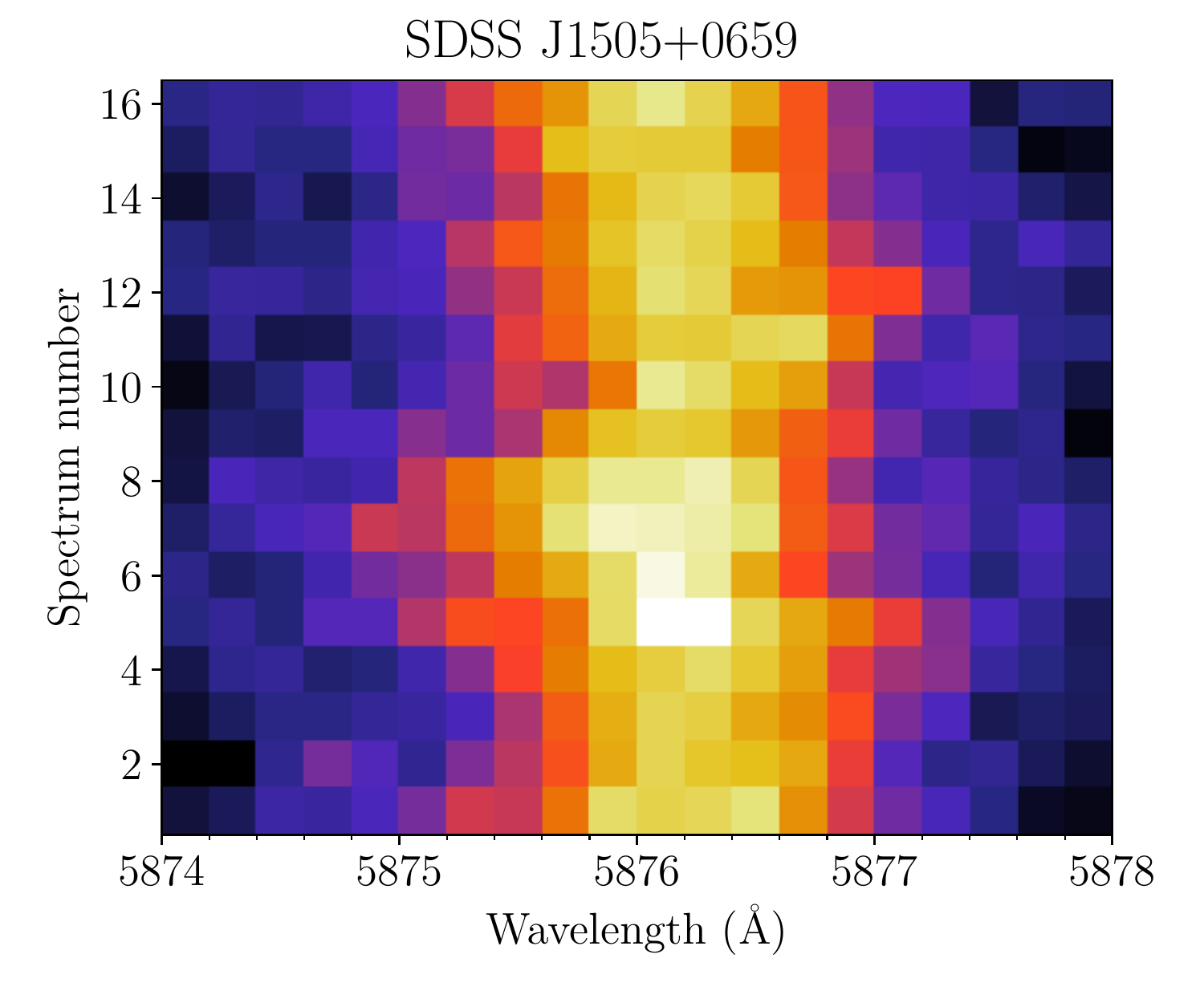}
\caption{Trailed spectra of the \hei\ 5875.6\,\AA\ line in SDSS\,J1505+0659, spanning a time period of 120\,min, and showing the velocity excursions of the central spike. Note that the spike is slightly redshifted from its rest wavelength, as is commonly seen.
}
\label{fig:j1505-trail}
\end{center}
\end{figure}

SDSS\,J1505+0659 is a known AM\,CVn binary discovered by \citet{Carter2014}. 
The system was discovered as part of a spectroscopic survey for AM\,CVn systems, selecting targets according to their SDSS colours \citep{Carter2013}.
The discovery paper included four spectra of the system retrieved from the SDSS database, from which they suggested an orbital period of 50.6\,min. 
The uncertainties on their measured radial velocities were large compared to the amplitude of the variation, due to the low resolution of SDSS spectra. Given this and the small number of data points available at the time, the orbital period was considered preliminary and subject to significant uncertainty.

\subsubsection{Spectroscopy}

Our combined VLT spectrum of SDSS\,J1505+0659 is shown in Figure~\ref{fig:avspec-xshooter}. 
The spectrum consists of a series of narrow helium emission lines imposed on a blue continuum. The profile of these helium emission lines is dominated by a strong central spike emission feature with relatively weak emission wings (Figure~\ref{fig:profiles-6}).
Central spike features are seen in many AM\,CVn binaries, and are emitted from on or near the surface of the accreting white dwarf \citep{Green2019}, though the emission mechanism is uncertain.
As was noted by \citet{Carter2014}, the strength of the central spike emission seen here is similar to a number of other long-period AM\,CVn binaries, including SDSS\,J1208+3550 \citep[orbital period 53.0\,min;][]{Kupfer2013}, SDSS\,J1137+4054 \citep[59.6\,min;][]{Carter2014}, and V396\,Hya \citep[65.1\,min;][]{Ruiz2001,Kupfer2016}.

Emission features are seen from nitrogen and sodium (we show the latter in Figure~\ref{fig:profiles-6}).
The presence of nitrogen emission is reminiscent of GP\,Com \citep[46.5\,min;][]{Nather1981,Marsh1990,Kupfer2016} and V396\,Hya \citep{Kupfer2016}. 
The \ion{Na}{I} emission is single-peaked and does not undergo any measurable velocity excursions. 
As such, we suggest it may share an origin with the \hei\ and \heii\ central spikes. (Given the weaker strength of the \ion{Na}{I} lines, it is unlikely that the small velocity excursions of the central spike, discussed below, would be detected for \ion{Na}{I}.)
It is difficult for neutral sodium and ionised helium to exist at the same temperature range, suggesting that the origin of the central spike emission must be sufficiently extended to include a range of temperatures.
In addition, we find a broad absorption feature centred on the \ion{Na}{I} 5890-95\,\AA\ doublet.
Absorption of this breadth is likely to be photospheric in origin.
This suggests a cool accreting white dwarf, as it is difficult to reconcile \ion{Na}{I} absorption with an atmosphere hotter than $\sim 10\,000$\,K.

The detection of sodium combined with a non-detection of calcium is unusual, and is worthy of note as it suggests that atmospheric sodium has been enhanced to levels higher than solar.
This may provide insight into the prior nature of the donor star.
Sodium can be produced via the Ne-Na cycle in the cores of high-mass main sequence stars \citep{Denisenkov1987} or in the helium-burning shells of asymptotic giant branch stars \citep{Mowlavi1999}.
On the other hand, it should be noted that isolated sodium has been seen in at least one accreting binary with an unevolved, brown dwarf donor \citep{Longstaff2019}.

The central spike feature undergoes periodic, low-amplitude velocity excursions, as shown in the trailed spectrum (Figure~\ref{fig:j1505-trail}).
These excursions were measured by cross-correlating each of the eight strongest emission lines in each spectrum with a Gaussian function. 
The radial velocities (RVs) measured in this way follow a sinusoidal modulation with a semi-amplitude of $5.0 \pm 0.2$\,km/s, around a systemic velocity ($\gamma$) which varies between different emission lines.
In other AM\,CVn binaries, central spike $\gamma$ values have been explained as a combination of an overall gravitational redshift and wavelength-dependent blueshifts arising from the Stark effect \citep{Morales-Rueda2003,Kupfer2016,Green2019}.
We performed an initial sinusoidal fit to the measured RVs for each line individually, in order to measure $\gamma$ for each line. 
The values of $\gamma$ measured in this way are shown in Table~\ref{tab:gammas}.
These $\gamma$ values were then subtracted from the RVs in order to perform a combined sinusoidal fit to all emission lines. 

\begin{table}
\caption{Measured systemic radial velocities for the \hei\ and \heii\ lines of SDSS\,J1505+0659. These systematic velocity shifts were measured using a sinusoidal fit to the measured radial velocities.}
\label{tab:gammas}
\begin{tabular}{cc}
\hline
Wavelength (\AA) & $\gamma$ (km/s)\\
\hline
\hei\  4471  & $ 39.5 \pm 1.0$\\
\heii\ 4686  & $ 25.1 \pm 1.1$\\
\hei\  4713  & $ 37.8 \pm 2.0$\\
\hei\  5015  & $ 21.3 \pm 1.6$\\
\hei\  5876  & $ 23.8 \pm 0.7$\\
\hei\  6678  & $ 27.5 \pm 1.3$\\
\hei\  7065  & $ 30.2 \pm 1.5$\\
\hei\  7281  & $ 28.9 \pm 2.0$\\
\hline
\end{tabular}
\end{table}

\begin{figure}
\begin{center}
\includegraphics[width=\columnwidth]{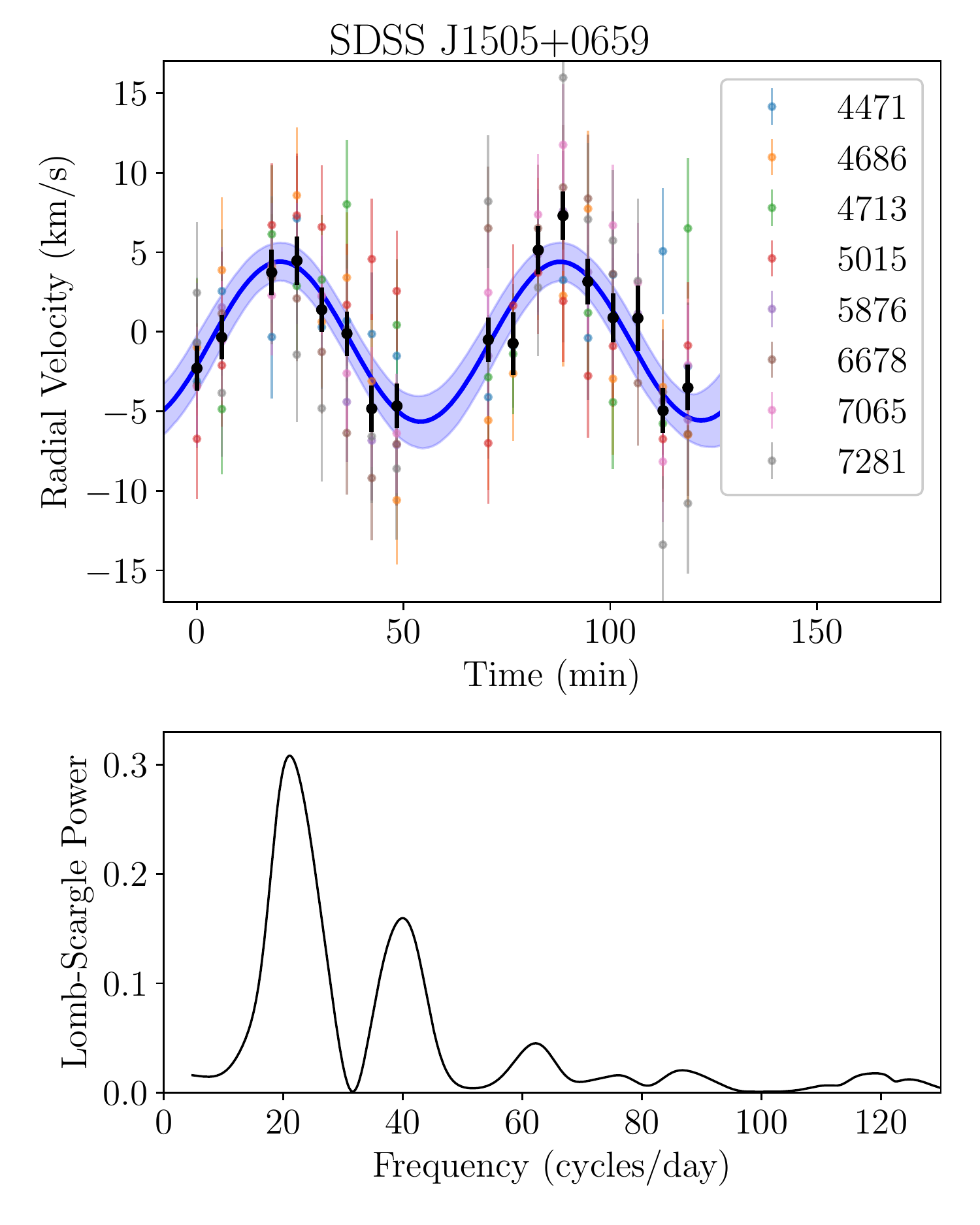}
\caption{\textit{Above:} Weighted mean RV measurements of SDSS\,J1505+0659 are shown as black points, while measurements from individual lines are shown as faded, coloured points. The median sinusoidal fit and 2-$\sigma$ range from a Markov Chain Monte-Carlo sampler is shown in blue, with a best-fit period of 67.8\,min. Uncertainties on RV measurements have been increased, as discussed in the text.
\textit{Below:} Lomb-Scargle plot produced from the measured RVs. 
}
\label{fig:j1505-sine}
\end{center}
\end{figure}


The $\gamma$-subtracted data and their best-fit sinusoid are shown in Figure~\ref{fig:j1505-sine}, along with a Lomb-Scargle periodogram \citep{Lomb1976,Scargle1982,VanderPlas2018}.
The sinusoidal fit was produced using a Markov Chain Monte Carlo method \citep[MCMC,][]{Goodman2010,Foreman-Mackey2013}.
The nominal uncertainties on individual RV measurements were underestimated, as evidenced by a reduced $\chi^2$ value of significantly more than one.
We compensated for this by combining the individual uncertainties in quadrature with an additional uncertainty term.
This term was included as a parameter in the MCMC, with a log-uniform prior.
In Figure~\ref{fig:j1505-sine} the uncertainties on the RV measurements have been scaled to include this additional uncertainty.
The orbital period produced by this process is $67.8 \pm 2.2$\,min.

This newly-measured period is significantly different from the 50.6\,min period suggested by \citet{Carter2014}. Given the uncertainties on their individual RV measurements, this discrepancy is not surprising.
This is the longest orbital period of any known AM\,CVn binary -- the longest previously published orbital period is that of V396\,Hya at $65.1 \pm 0.7$\,min \citep{Ruiz2001}.


When compared to the empirical correlation between orbital period and emission line strength noted by \citet{Carter2013}, SDSS\,J1505+0659 is a clear outlier. For the \hei\ 5876\,\AA\ line, we measure an equivalent width (EW) of $-37.9 \pm 3.2$\,\AA. For a system with an orbital period of 68\,min, Carter's relation would predict an EW of approximately $-90$\,\AA. 
However, there are several other systems with smaller EWs than would be predicted by the correlation, including SDSS\,J1208+3550 and SDSS\,J1137+4054.
The unexpectedly weak EW of SDSS\,J1505+0659 may suggest that the disc is unusually weak.
There is also likely to be an inclination dependence in the EW which may cause systems to be outliers.


\begin{figure}
\begin{center}
\includegraphics[width=\columnwidth]{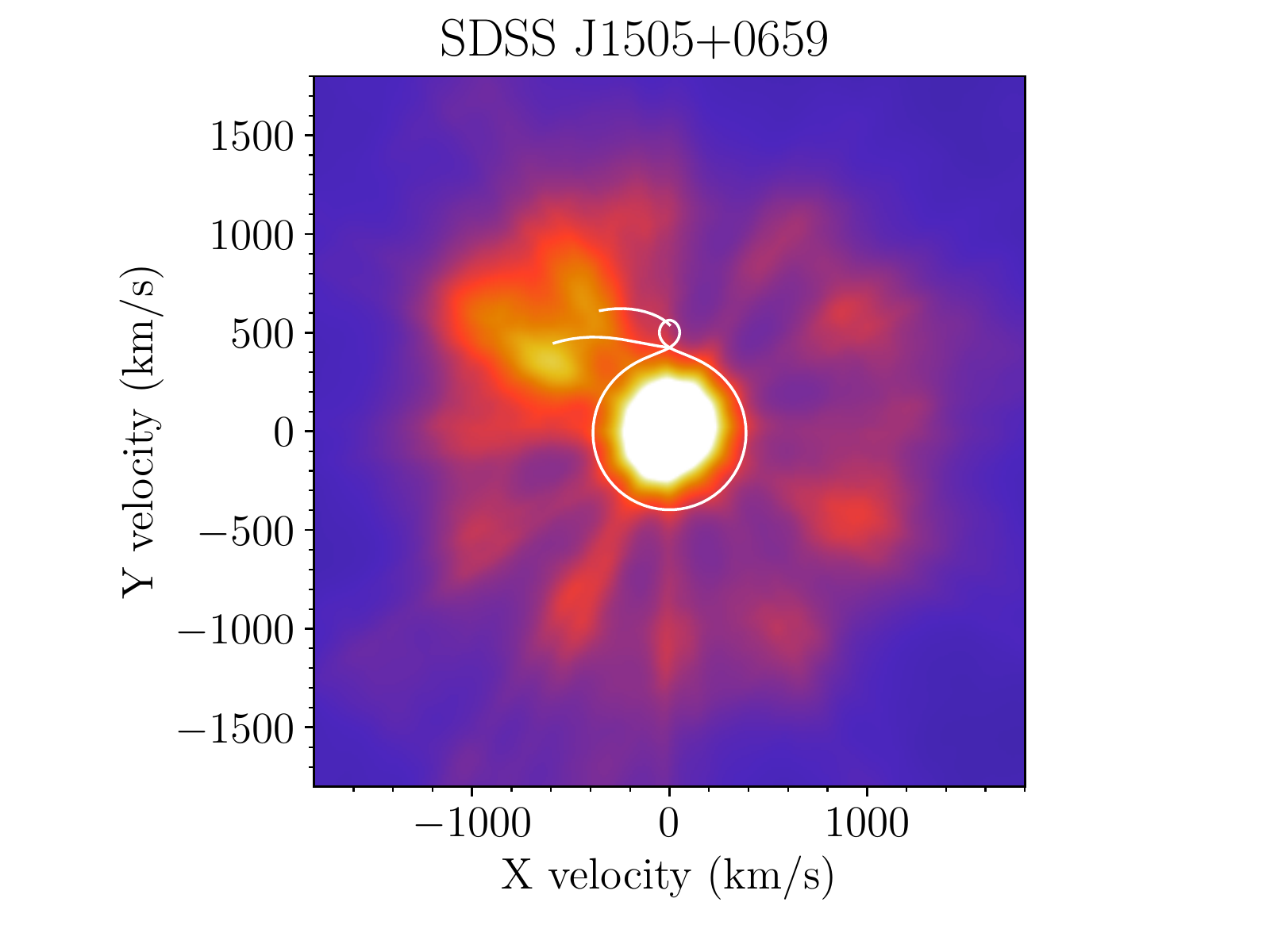}
\caption{Doppler map of SDSS\,J1505+0659 produced from the 5015\,\AA\ \hei\ line, and rotated to the standard orientation based on the RV fit in Figure~\ref{fig:j1505-sine}. 
Emission is dominated by the bright central spike, but an accretion disc and a bright spot can also be seen. 
Radial spokes are an artefact of poor phase coverage.
The superimposed white lines show predicted positions of various components, including: the Roche lobes of the two stars \viva{(the lobe of the accretor centred approximately on zero, and the donor star above this, centred on approximately $V_y = 450$\,km/s)}; the velocities of a ballistic stream of infalling matter; and the Keplerian velocities at each point along the stream. The latter two are shown at positions beginning from the inner Lagrange point and ending at a radius of 0.7 times the Roche lobe radius.
}
\label{fig:j1505-dopp}
\end{center}
\end{figure}

\subsubsection{Mass Ratio}

The mass ratio can be calculated from a combination of other binary properties: the orbital period ($P_\mathrm{orb}$), the velocity amplitude ($K_1$), the orbital inclination ($i$), and the mass of the accretor ($M_1$). 
The calculation uses Kepler's law and the necessity that orbital velocity $v_1 = K_1/\sin i = 2 \pi r / P_\mathrm{orb}$, where $r$ is the separation of the accreting star from the centre of mass of the system.
We have measurements of $\porb = 67.8 \pm 2.2$\,min and $K_1 = 5.0 \pm 0.2$\,km/s from the sinusoidal fit shown in Figure~\ref{fig:j1505-sine}.
For $M_1$, we assumed $0.8 \pm 0.1 M_\odot$. This value is typical of CV accretors \citep{Pala2020} and consistent with the two AM\,CVn accretors which have measured masses \citep{Copperwheat2011,Green2018}.

In order to constrain the orbital inclination, we measured the velocity of the peak intensity of the disc-originating emission line wings (ie.\ the outer two peaks of the triple-peaked line profile). The peak intensity occurred at $v_D = 750 \pm 250$\,km/s. 
This peak emission originates from close to the edge of the accretion disc, at a radius described by \citep{Smak1981,Warner1995}
\begin{equation}
v_D^2 = \frac{M_1 G \sin^2(i)}{r_d (0.95 \pm 0.05)^2}.
\end{equation}
In this equation $r_d$ is the disc radius and $G$ is the gravitational constant. We assume that the disc radius is equal to 0.7 times the radius of the Roche lobe of the accretor, as is often seen in CVs \citep{Warner1995} and in the eclipsing AM\,CVn, Gaia14aae \citep{Green2018}.
The Roche lobe radius as a fraction of the orbital separation of the two stars ($a$) can be calculated as \citep{Eggleton1983}
\begin{equation}
\frac{R_L}{a} = \frac{0.49 q^{-2/3}}{0.6 q^{-2/3} + \ln (1 + q^{-1/3})}.
\label{eq:rlobe}
\end{equation}

We combined these constraints using a MCMC routine. 
The median result was $q = 0.011 ^{+ 0.003} _{- 0.001}$, where the uncertainties refer to the 16th and 84th percentiles of the distribution. The corresponding inclination was $i = 60 \pm 20 ^\circ$.
This mass ratio is similar to that of V396\,Hya, the previously longest-period AM\,CVn binary, which has $q = 0.014 \pm 0.004$ \citep{Kupfer2016}.

A Doppler map of the system was produced using the 67.8\,min orbital period. The \hei\ 5015\,\AA\ line was chosen because it has prominent disc emission. The Doppler map is shown in Figure~\ref{fig:j1505-dopp}. The quality of this map is limited by the small number of spectra, giving the map spoke-like artefacts.
Despite this, several features can be distinguished. In addition to the central spike and disc emission, a bright spot can be seen in the top left quadrant. As this is the standard location for the bright-spot in Doppler maps, it provides further support to the RV measurements of the central spike which define the orbital phase.
Figure~\ref{fig:j1505-dopp} also shows predicted velocity tracks for the infalling matter stream, based on the values from our MCMC fit, which approximately coincide with the bright spot.

\subsubsection{Photometry}

The long-term lightcurve of SDSS\,J1505+0659 (Figure~\ref{fig:longterm-lc}) shows no activity over the period monitored. 
There are several ASAS-SN detections that are brighter than the baseline. However, as each is an isolated detection preceded and followed by non-detections, and consistent with the magnitude limit of ASAS-SN, we believe these to be false detections.
For an AM\,CVn of this orbital period, outbursts are not expected due to the low accretion rate.

In high-speed photometry of the system we detect no significant variability. 
By injecting a sinusoidal signal at the orbital period, and assuming that we would detect any signal that had a Lomb-Scargle peak at four times the mean Lomb-Scargle power, we estimate an upper limit on any orbital signal amplitude at 0.5 per cent of the system brightness.

\begin{figure}
\begin{center}
\includegraphics[width=\columnwidth]{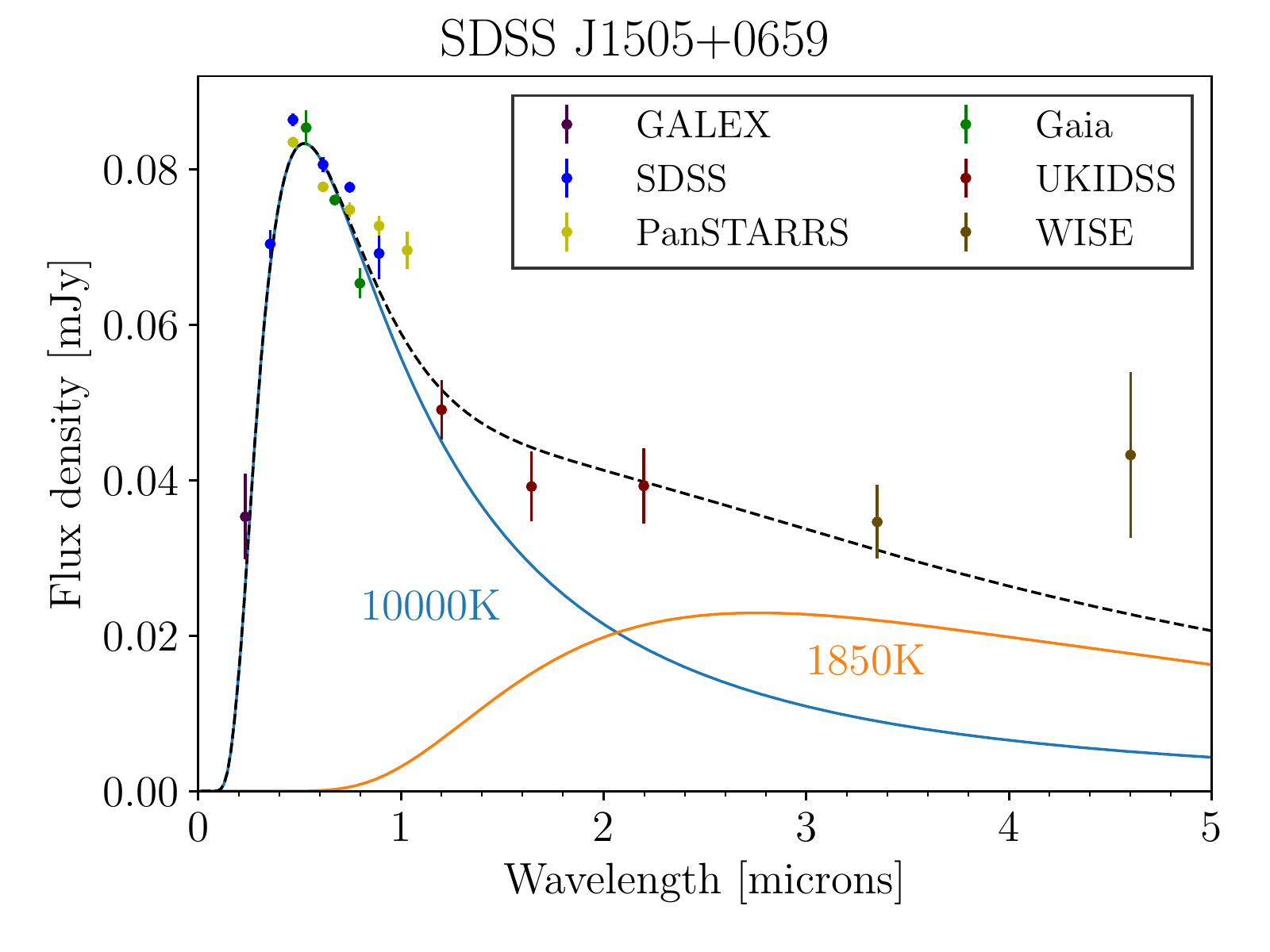}
\caption{Spectral energy distribution of SDSS\,J1505+0659. The blue curve shows a blackbody spectrum at a temperature estimated from the optical fluxes, corrected for the predicted amount of reddening. The orange curve shows a second blackbody spectrum with radius fixed to the Roche lobe radius of the donor star, and the black dashed curve shows the sum of the two blackbody spectra. 
The addition of the second blackbody explains the infrared excess seen in the UKIDSS and WISE data.
}
\label{fig:j1505-sed}
\end{center}
\end{figure}

The spectral energy distribution (SED) of SDSS\,J1505+0659 is noteworthy. 
In Figure~\ref{fig:j1505-sed} we show flux densities measured by 
SDSS \citep{Abazajian2009},
\textit{Gaia} \citep{GaiaCollaboration2018},
Pan-STARRS \citep{Chambers2016},
the Galaxy Evolution Explorer \citep[GALEX,][]{Morrissey2007},
the UKIRT Infrared Deep Sky Survey \citep[UKIDSS,][]{Lawrence2007},
and the Wide-field Infrared Survey Explorer \citep[WISE,][]{Wright2010}.
We show a blackbody curve of 10\,000\,K, corrected for the expected amount of reddening and arbitrarily scaled to match the optical data.
An infrared excess can be seen in both UKIDSS and WISE data.
As the helium-atmosphere accreting white dwarf should be well approximated by the blackbody, and the accretion disc should be of a similar temperature, it is likely that the infrared excess comes from the donor star.
We also plot a second blackbody, with a radius equal to the Roche lobe radius predicted by Equation~\ref{eq:rlobe} and a temperature of 1850\,K (chosen to fit the infrared data).
At this temperature the donor would be approximately L4 spectral type.
The sum of the two blackbodies is a reasonable fit for all data.
If our interpretation is correct, it is the first published direct detection of light from the donor star in a non-direct impacting AM\,CVn binary.
However, while we consider the above to be the most likely interpretation, we caution that some non-thermal component originating at the accretion disc could also result in this SED.

\subsection{ASASSN-14ei}
\label{sec:14ei}

\begin{figure*}
\begin{center}
\includegraphics[width=500pt]{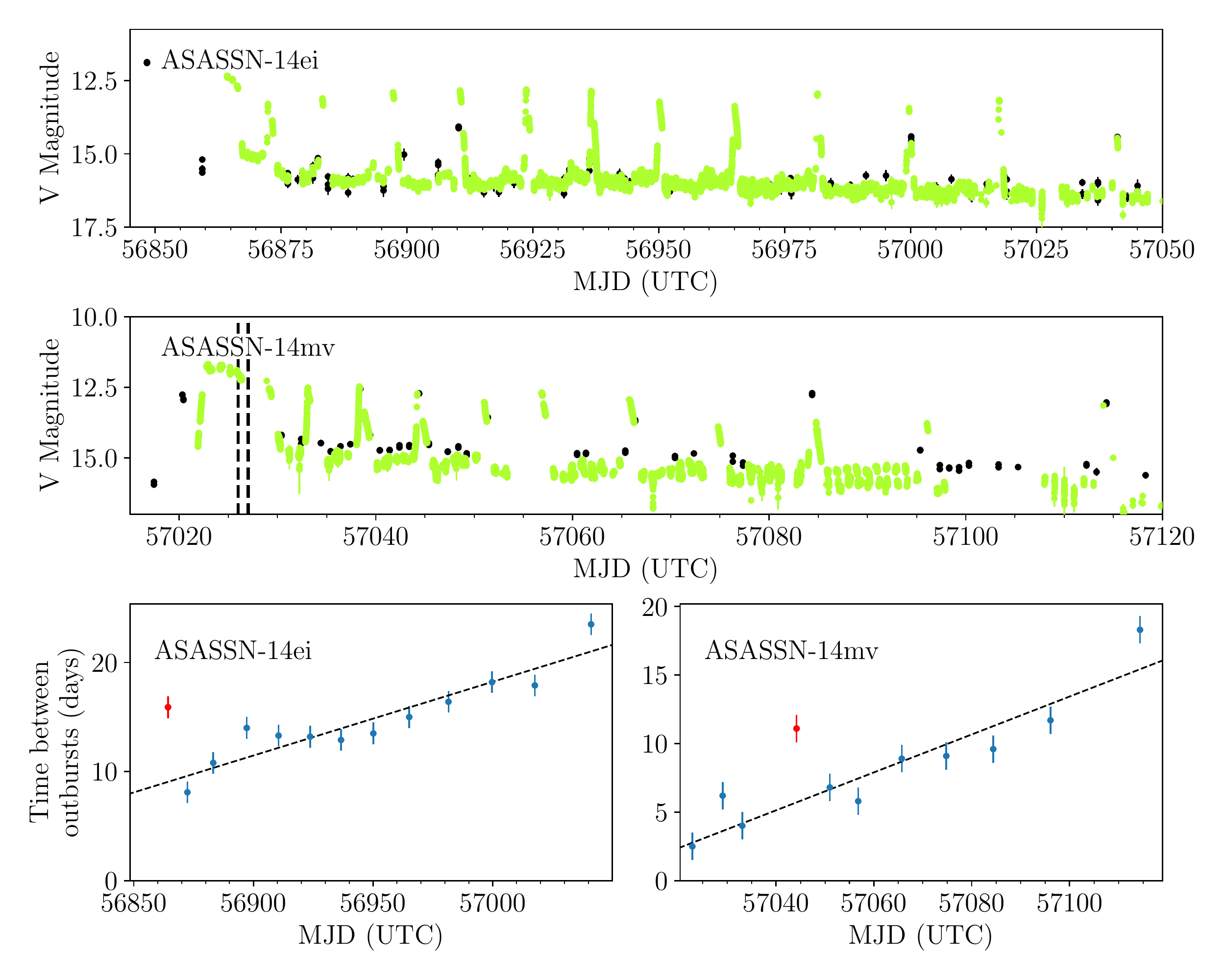}
\caption{\textit{Top} and \textit{middle}: Echo outburst lightcurves of ASASSN-14ei and ASASSN-14mv, showing the echo outburst series observed in 2014. Data are coloured according to the scheme in Figure~\ref{fig:longterm-lc}, with green points from the AAVSO and black points from ASAS-SN..
Vertical dashed lines show the dates of ULTRASPEC observations of ASASSN-14mv.
\textit{Bottom left} and \textit{right}: The time separation between the peak of each detected echo outburst and the peak of its preceding echo outburst.
Blue points were used to produce a linear fit (dashed line) while red points were considered outliers and were excluded.
The peak of each outburst was selected as the time of the brightest measured AAVSO detection within that outburst.
Uncertainties were chosen as $\pm 1$\,day, to account for sampling limitations arising from the day/night cycle.
In both cases the origin of the x-axis is aligned with the time of the original ASAS-SN detection.
}
\label{fig:echos}
\end{center}
\end{figure*}

\begin{figure}
\begin{center}
\includegraphics[width=\columnwidth]{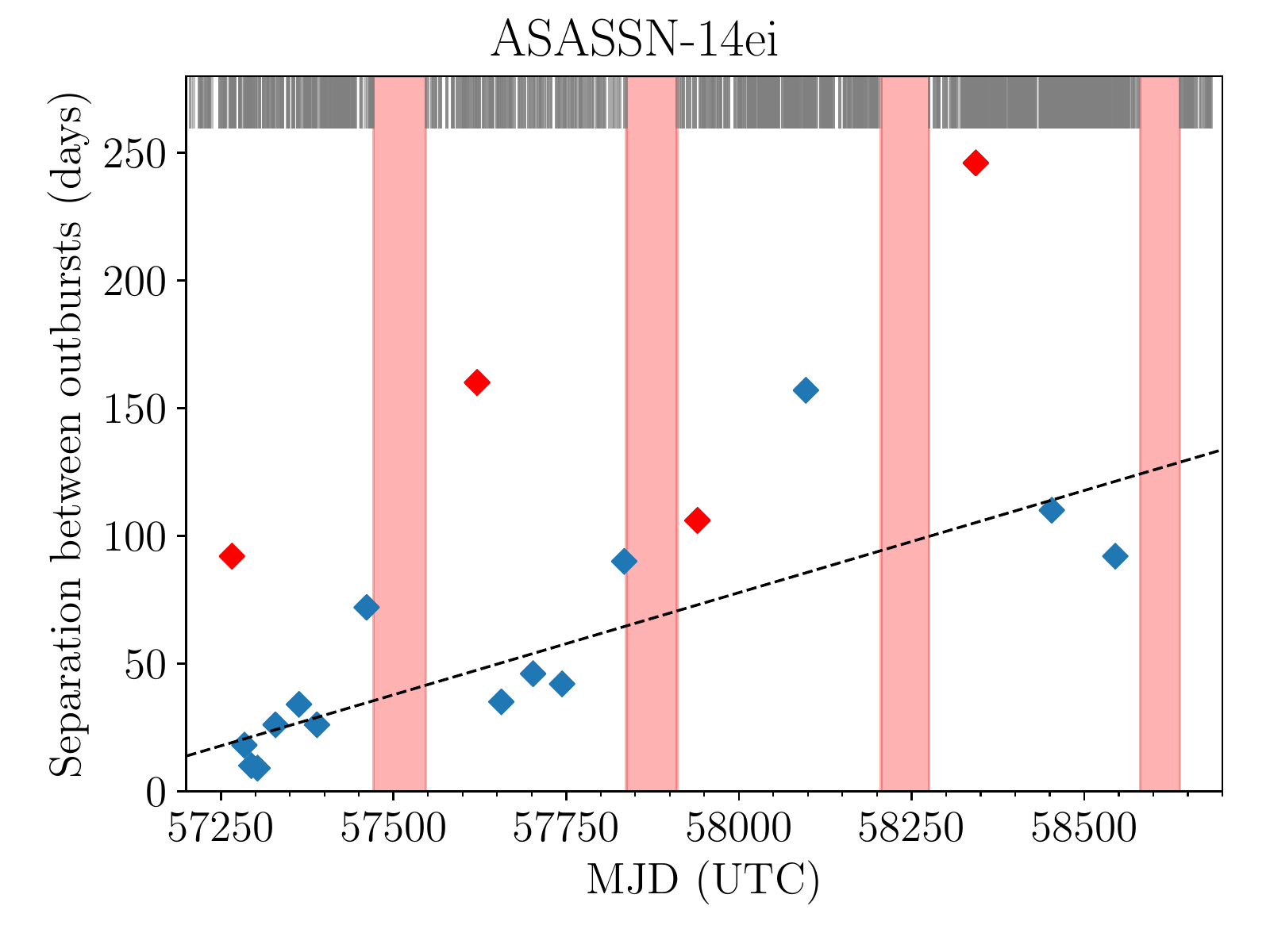}
\caption{Separations between the times of each normal outburst and its preceding outburst for ASASSN-14ei are shown as blue points. 
This figure shows isolated outbursts, not including echo outbursts, which were shown in Figure~\ref{fig:echos}.
All outburst detections shown here are from ASAS-SN or AAVSO. 
Grey vertical dashes at the top of the plot denote the times of ASAS-SN observations; there is no significant decrease in cadence to explain the growing separations of detected outbursts. 
Red highlighted regions denote significant gaps in coverage due to the star being behind the sun. The data point following each gap has been coloured red to show that it may be unreliable (as preceding outbursts may have been missed during the gap).
The dashed black line shows a linear least-squares fit to all blue points, highlighting the trend towards increasing intervals between outbursts.
}
\label{fig:14ei-outbursts}
\end{center}
\end{figure}

ASASSN-14ei (also known as OX Eri) is an AM\,CVn binary, first detected in superoutburst by ASAS-SN in 2014 July \citep{Prieto2014}.
Follow-up observations reported \review{by} \citet{Prieto2014b} include a spectrum showing \hei\ and \heii\ lines, as well as optical, UV and X-ray imaging with \textit{Swift}, and a series of rebrightenings following the original outburst.
VSNET alert \#17575\footnote{http://ooruri.kusastro.kyoto-u.ac.jp/mailarchive/vsnet-alert/17575} reported a photometric period during outburst of 41.63(3)\,min, believed to be a superhump period. Based on archival data from the CRTS, they suggest a photometric period of $\approx 40.0$\,min during quiescence, possibly the orbital period.

\subsubsection{Long-term Lightcurve}

The long-term lightcurve of ASASSN-14ei is shown in Figure~\ref{fig:longterm-lc}. 
There are several unusual features in the long-term lightcurve of ASASSN-14ei.
No outburst activity is detected during the CRTS coverage before the 2014 superoutburst.
Then, following the original superoutburst in 2014, a series of 12 echo outbursts were seen and were followed by the AAVSO. 
Since the end of the echo outbursts, a number of standard outbursts have been seen with a recurrence time of less than 100\,days, shorter than is generally seen in AM\,CVn binaries.
In addition, since the superoutburst, the magnitude of the system between outbursts (in both ASAS-SN and AAVSO data) is brighter than the CRTS magnitude before superoutburst by around 1\,mag.
We will discuss each of these features in turn.

The series of 12 echo outbursts following the 2014 superoutburst were observed intensively by the AAVSO.
The lightcurve of these outbursts is shown in Figure~\ref{fig:echos}.
Such a large number of echo outbursts is highly unusual, having been seen in only one previous system, WZ\,Sge \citep{Patterson2002}.
It can be seen in the top panel of Figure~\ref{fig:echos} that the time separation of these echo outbursts increases over time.
In the bottom right panel of Figure~\ref{fig:echos} we show the separation between each measured echo outburst. The separation between outbursts increases appoximately linearly with the elapsed time since the original superoutburst.

During these echo outbursts, a roughly sinusoidal variation in brightness is seen.
In order to measure the period of this signal, we divided the available AAVSO data by observer and then by night. 
For each of the resulting sections, a Lomb-Scargle periodogram was produced.
From all of the periodograms produced in this way, we selected those in which the signal was visible. The mean period from these nights was $42.85 \pm 0.14$\,min, where the uncertainty is the standard error on the mean ($\sigma / \sqrt{N}$).

Following the 2014 superoutburst, the system appears quite different to its state before the superoutburst, being around 1\,mag brighter (though note that CRTS uses a clear filter whereas ASAS-SN uses $V$-band and $g'$-band filters) and showing frequent, normal outbursts (ie.\ not superoutbursts) which were not seen before 2014. 
The increase in magnitude can be explained by heating of the central white dwarf. However, the increase in outburst rate is harder to explain.
One possible explanation is that the 2014 superoutburst changed the state of the system in such a way that the accretion rate has increased.
Several sources have previously suggested that mass transfer rate in hydrogen-dominated CVs may increase via irradiation of the donor, and the phenomenon was invoked by \citet{Kotko2012} to explain various properties of AM\,CVn superoutburst lightcurves. 
However, modelling of the irradiation process raises several concerns, most notably the shielding of the equator of the donor star by the accretion disc, and the difficulty of transferring any heating from the higher latitudes of the donor to the equator \citep{Osaki2003,Osaki2004}.

Following this perturbation to the state of the system, there is evidence that ASASSN-14ei is slowly returning to the quiescent state that it was in before the 2014 superoutburst.
The magnitude of the system appears to be fading. Based on the ASAS-SN $V$-band data (spanning 1300 days, the longest available stretch of data with a consistent observing setup), after masking all outbursts, the brightness is decreasing at a rate of $0.8 \pm 0.2$\,mag/yr. 
The frequency of outbursts is also decreasing. In Figure~\ref{fig:14ei-outbursts}, we show the separation between normal (ie.\ non-echo) outbursts as a function of time. While there is considerable scatter, the overall trend appears to be that outbursts are becoming less frequent.
Both of these changes would be consistent with a system returning towards its quiescent state after being perturbed to a higher accretion rate during superoutburst.

The above explains an unusual observation noted by \citet{Ramsay2018}. 
That paper found ASASSN-14ei to be unusually bright for its orbital period, with an absolute magnitude $M_g = 8.94$.
This absolute magnitude was based on an apparent magnitude from SkyMapper of $g = 15.5$.
It can be seen in Figure~\ref{fig:longterm-lc} that this is brighter than the quiescent magnitude of ASASSN-14ei.
We can explain this by noting that the reported epoch of the SkyMapper magnitude measurement is MJD\,56963, which was just after the 2014 superoutburst when the brightness of the system was still significantly elevated.
A more reliable estimate of the quiescent magnitude can be taken from the CRTS observations before the superoutburst, which have a mean (unfiltered) apparent magnitude of 18.06\,mag. This brings the absolute magnitude to $M = 11.5$\,mag, very similar to other AM\,CVn binaries of the same orbital period.

\subsubsection{High-speed Photometry}

\begin{figure}
\begin{center}
\includegraphics[width=\columnwidth]{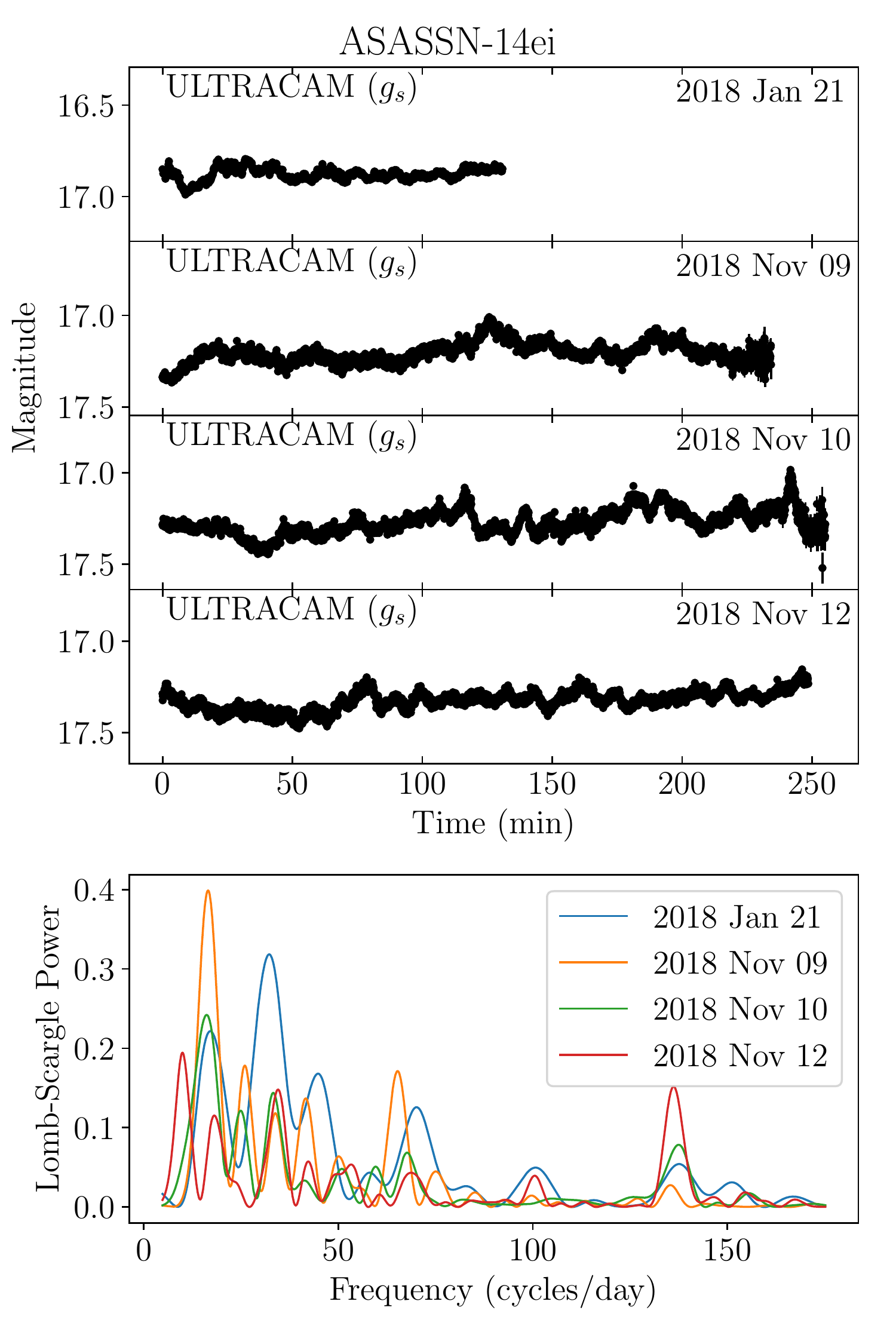}
\caption{\textit{Upper panels:} ULTRACAM photometry of ASASSN-14ei from four nights. \textit{Lower panel:} Lomb-Scargle periodograms of the data shown in the upper panels, after first subtracting a three-term polynomial.
}
\label{fig:14ei-phot}
\end{center}
\end{figure}

We obtained high-speed photometry of ASASSN-14ei with ULTRACAM across four nights, one in 2018 January and three in 2018 November (Figure~\ref{fig:14ei-phot}). 
The system was not in outburst for any of these observations, although a difference in magnitude can be seen between January and November (consistent with the decreasing trend in brightness noted in the previous section).

On January 21, the strongest signal has a period of 44.6\,min.
Across two consecutive nights, November 9 and November 10, periods can be seen at 86.9\,min and 88.7\,min respectively. 
On November 12, this signal is not seen.
It may be that the signal from January 21 is a harmonic of the signal seen on November 9 and 10, as the periodogram does show a smaller peak at around this frequency.
We do not know of any physical process that would produce a signal in the 80-90\,min range, but note that AM\,CVn binaries have been known to show quasi-periodic oscillations at a variety of frequencies \citep{Fontaine2011}.
None of these signals is consistent with the periods discussed previously in this paper: neither the period we measured from the AAVSO outburst data ($42.85 \pm 0.14$\,min), nor the period measured by VSNET during outburst ($41.63 \pm 0.03$\,min), nor the suggested period during quiescence from CRTS ($\approx 40.0$\,min).

We also note that a signal with a period of approximately 10.5\,min appears to be present in data from all four nights. 
Such a signal could perhaps be related to the spin period of the central white dwarf, but may also be a quasi-periodic oscillation.

\subsubsection{Spectroscopy}

\begin{figure}
\begin{center}
\includegraphics[width=\columnwidth]{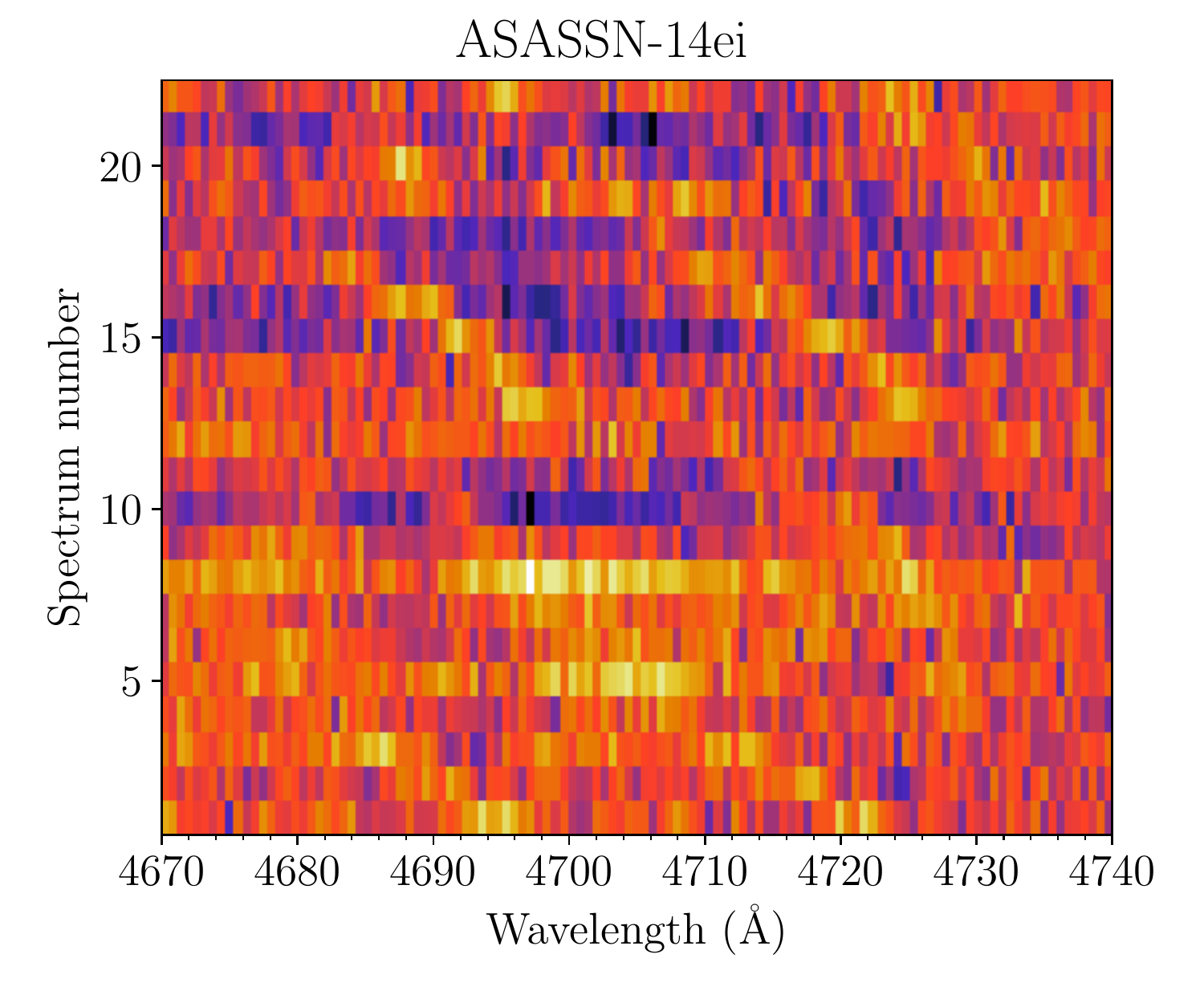}
\caption{Trailed spectra of the \heii\ 4686\,\AA\ and \hei\ 4713\,\AA\ lines in ASASSN-14ei, spanning an observation window of 120\,min. Each spectrum has been mean-subtracted. At some phases, weak S-waves can be seen in both lines (visible as signals moving in parallel) resulting from the bright spot of the system.
}
\label{fig:14ei-trail}
\end{center}
\end{figure}

\begin{figure}
\begin{center}
\includegraphics[width=\columnwidth]{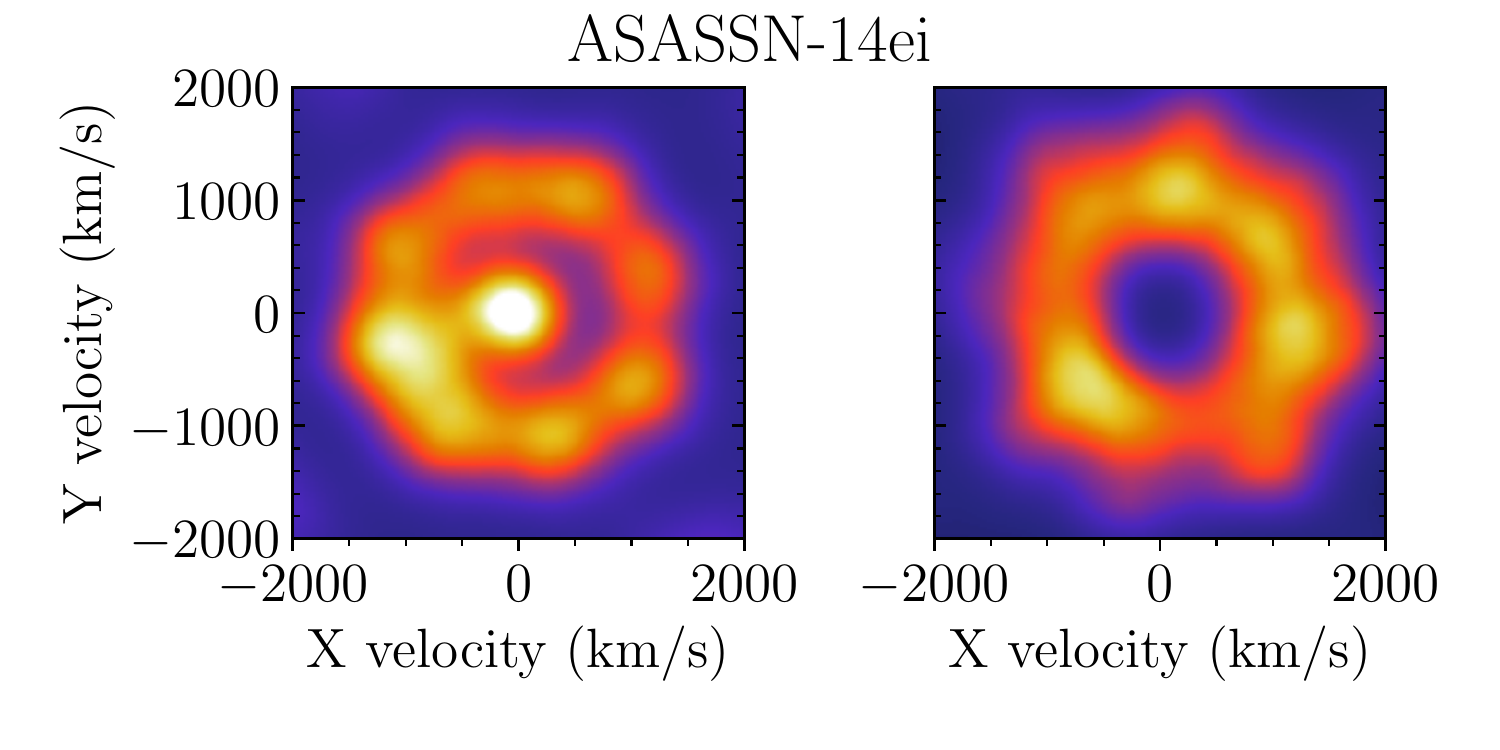}
\caption{Doppler maps of ASASSN-14ei produced simultaneously from the 4686\,\AA\ \heii\ line \textit{(left)} and the 4713\,\AA\ \hei\ line \textit{(right)}. 
Regularly spaced artefacts around the accretion disc result from aliasing between the observation times of the spectra and the orbital period.
The \heii\ line features a strong central spike, an accretion disc, and possibly a bright spot on the left hand side.
The \hei\ line shows accretion disc emission, but no central spike. There may be also be a bright spot on the left hand side.
}
\label{fig:14ei-dopp}
\end{center}
\end{figure}

The mean X-shooter spectrum of ASASSN-14ei is shown in Figure~\ref{fig:avspec-xshooter}. A series of helium emission lines are seen, including both \hei\ and \heii. Several metal lines are also seen, including calcium, nitrogen, and silicon in emission and magnesium and sodium in absorption. 
The nitrogen lines are notably weaker than in SDSS\,J1505+0659.
The presence of magnesium absorption is interesting. A trailed spectrum shows no notable velocity excursions from the magnesium triplet, suggesting that it originates from the central white dwarf. This may imply a low temperature for the central white dwarf, given the low ionisation energy of magnesium.
If the mass transfer rate of the system is indeed elevated as suggested in the previous section, this low temperature would be unexpected.

Trailed spectra of some helium lines show marginally-visible S-waves from the bright spot (Figure~\ref{fig:14ei-trail}).
Each individual spectrum has been continuum- and mean-subtracted, in order to remove stationary components (including the accretion disc) and highlight the velocity-varying bright spot.
The period of these S-waves is roughly consistent with the orbital period of $\approx 40$\,min proposed previously.
Unfortunately the bright spot is not strong enough to be detected by the cross-correlation method used for SDSS\,J1505+0659.

In order to explore the orbital period, Doppler maps were produced for a range of periods between 38 and 43\,min. 
Each map was inspected visually in order to determine those maps in which the bright spot was most clearly visible. 
Such a qualitative approach is clearly not ideal; however, aliasing between the cadence of observations and the orbital period produces artefacts in the maps which are likely to confound quantitative measures such as entropy \citep{Wang2017,Wang2018}.
Our preferred maps had orbital periods in the range of 41 to 42.5\,min.
In Figure~\ref{fig:14ei-dopp} are shown Doppler maps produced using a period of 41.6\,min.

The orbital periods preferred by this Doppler tomography method are slightly longer than the CRTS photometric orbital period of 40.0\,min, and agree better with the proposed superhump period from VSNET of 41.63(3)\,min. It may be that the modulation reported in VSNET was in fact orbital in origin. 
\review{Photometric signals} at the orbital period can be seen in some systems during some stages of the outburst \citep[eg.][]{Isogai2019}. 
However, with the current data we are unable to select between these possible orbital periods.


\subsection{ASASSN-14mv}

ASASSN-14mv (also known as V493 Gem) was discovered in superoutburst by ASAS-SN in 2014 December \citep{Denisenko2014}. 
These authors also discussed archival outburst detections from 1938 March (via the Heidelberg Astronomical Plates), 1991 March (via the Palomar Observatory Sky Survey II), and 2011 January (via the MASTER robotic network).
The 2011 outburst is also visible in CRTS and PanSTARRS data (see Figure~\ref{fig:longterm-lc}).
During the 2014-15 outburst, photometric periods of 41.8\,min (Stage A) and 41.0\,min (Stage B) were reported by VSNET alert \#18124\footnote{http://ooruri.kusastro.kyoto-u.ac.jp/mailarchive/vsnet-alert/18124}.

\subsubsection{Long-term Lightcurve}
\label{sec:14mv-ll}

The long-term lightcurve of ASASSN-14mv is reproduced in Figure~\ref{fig:longterm-lc}. 
Note that the data from ASAS-SN are affected by a systematic offset of 2.5\,mag.
Data from other sources (CRTS, PanSTARRS, AAVSO, \textit{Gaia}, and ULTRASPEC, many of these concurrent with the ASAS-SN data) all agree with each other that the quiescent magnitude of the binary is around 18th magnitude, while ASAS-SN measures the quiescent magnitude as 15.5.
For several other stars within 20", we compared the ASAS-SN and \textit{Gaia} magnitudes, and found the same 2.5\,mag offset. 
We therefore conclude that some unknown systematic problem is affecting ASAS-SN measurements of stars in this field.
Given that there are seven stars within 11" of ASASSN-14mv, and the pixel size of ASAS-SN is 8", it is possible that poor seeing may cause some contamination from nearby stars.
We therefore discount the ASAS-SN data of this target.

Like ASASSN-14ei, ASASSN-14mv underwent a series of echo outbursts. In this case, 10 rebrightenings followed the original superoutburst.
These outbursts were followed by the AAVSO, whose data are reproduced in Figure~\ref{fig:echos}.
As we noted for ASASSN-14ei, the separation between these echo outbursts increases over time, following an approximately linear trend when plotted against the total time elapsed since the original superoutburst.
The trend is not strict, however. The fourth rebrightening (marked in red in the bottom right panel of Figure~\ref{fig:echos}) comes approximately twice as late as expected. The coverage by AAVSO makes it clear that no rebrightening was missed between the third and fourth, meaning that this delay is genuine. Apart from this outlier, the trend is followed remarkably well.

No outbursts of ASASSN-14mv have been detected since the 2014 superoutburst and its rebrightenings, in stark contrast with ASASSN-14ei.
Without reliable ASAS-SN measurements, we have limited constraints on how quickly the system returned to its quiescent magnitude. 
We can say that it had returned to its quiescent magnitude by the time of the ULTRASPEC observation in 2017, some 1000 days following the 2014 superoutburst. 
(Although the filters are different, the 2017 ULTRASPEC \textit{KG5} observations are a similar magnitude to quiescent observations in the Pan-STARRS \textit{g}- and \textit{r}-bands, which cover a similar wavelength range.)
Following the previous outburst in 2011, we can say that the system returned to its quiescent magnitude after at most 320 days, based on CRTS observations.
This behaviour can also be contrasted with ASASSN-14ei, which has remained at an elevated brightness for over 1500 days.
Given that the systems appear similar in other ways (similar orbital periods, and both show a similar pattern of rebrightenings immediately following superoutburst), these differences in long-term behaviour is interesting to note.

We searched the AAVSO data during outburst for periodic signals following the methodology described for ASASSN-14ei.
The signals we detect have a mean value of $40.9 \pm 0.3$\,min, which is consistent with the superhump modulation reported by VSNET.

\subsubsection{High-Speed Photometry}

\begin{figure}
\begin{center}
\includegraphics[width=\columnwidth]{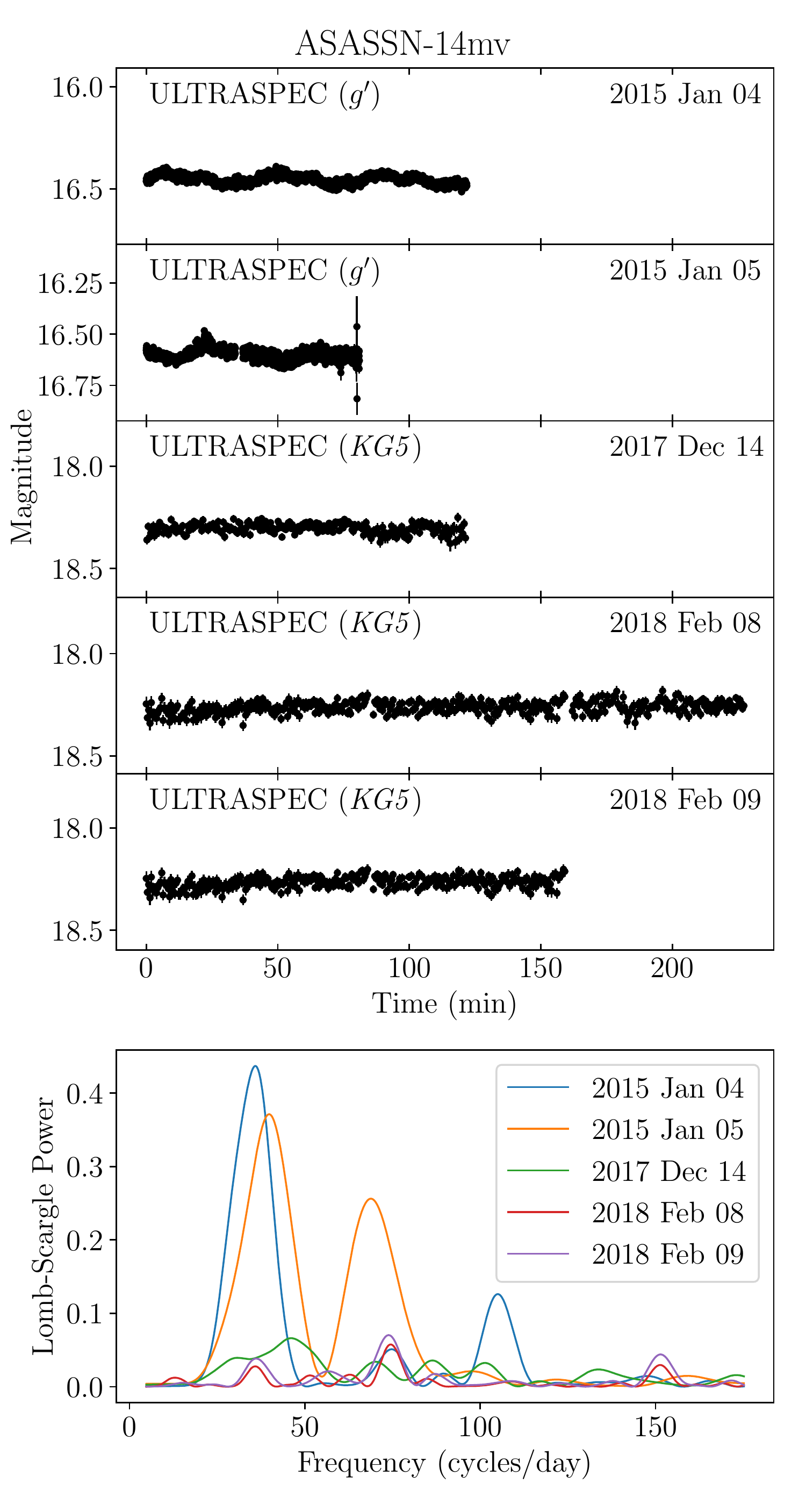}
\caption{\textit{Upper panels:} ULTRASPEC photometry of ASASSN-14mv from five nights. The 2015 data were obtained during superoutburst, whereas the 2017 and 2018 data were obtained during quiescence. \textit{Lower panel:} Lomb-Scargle periodograms of the data shown in the upper panels, after first subtracting a three-term polynomial.
}
\label{fig:14mv-phot}
\end{center}
\end{figure}

We observed ASASSN-14mv using ULTRASPEC on two consecutive nights during the 2014-15 superoutburst, and three nights in quiescence in 2017-18. 
These data are shown in Figure~\ref{fig:14mv-phot}.

During outburst, strong modulations are seen on both nights, with periods of 40.2\,min and 36.2\,min.
During quiescence, on two consecutive nights in 2018 February, a weaker but significant signal can be seen with periods of 40.2\,min and 39.8\,min. 
Apart from the January 05 signal, these periods are all consistent with the period detected in the AAVSO data. The January 05 signal has a slightly shorter period, but given the relatively short timespan of the observations on that night this may not be a significant difference.
As this modulation was most strongly seen in outburst, it is likely to be a superhump signal. It is consistent with the Stage~B superhump period reported by VSNET.

\subsubsection{Spectroscopy}

The mean spectrum of ASASSN-14mv is shown in Figure~\ref{fig:avspec-14mv}.
It shows a typical AM\,CVn spectrum consisting of a series of \hei\ and metal emission lines imposed on a blue continuum. 
The metals present include calcium, sodium, and nitrogen.
The nitrogen emission is relatively strong when compared to ASASSN-14ei.
The line profiles are narrow compared to other systems (Figure~\ref{fig:profiles-6}), suggesting a relatively face-on orbital inclination.
Due to the long exposure times used, the orbital RV shifts of the binary are not measurable.


\subsection{MOA\,2010-BLG-087}

MOA\,2010-BLG-087 was detected during outburst in 2010 by the Microlensing Objects in Astrophysics (MOA) survey of the Galactic bulge.
Follow-up photometry showed variability on a period of approximately 50 minutes, and a spectrum during outburst showed a series of helium lines, suggesting an AM\,CVn nature for the object (K.~Horne, priv.~comm.).
The target is in a crowded region of sky, complicating follow-up. Figure~\ref{fig:moa-finder} shows a finder chart, produced from a clipped-mean combination of ULTRACAM images with ~1" seeing. Note that these images were taken during outburst, and so the target was unusually bright at this time.

\subsubsection{Spectroscopy}

The spectrum of MOA\,2010-BLG-087 (Figure~\ref{fig:avspec-fors2}) is not typical for an AM\,CVn binary. Instead it appears similar to a K-type star.
We have visually inspected the acquisition images taken prior to these data, and are reasonably confident that the correct target was observed.
We suggest that the target is blended with a nearby K star in all ground-based observations to date. During outburst the AM\,CVn binary was the dominant source, while in quiescence the nearby star dominated.

\subsubsection{High-Speed Photometry}

\begin{figure}
\begin{center}
\includegraphics[width=\columnwidth]{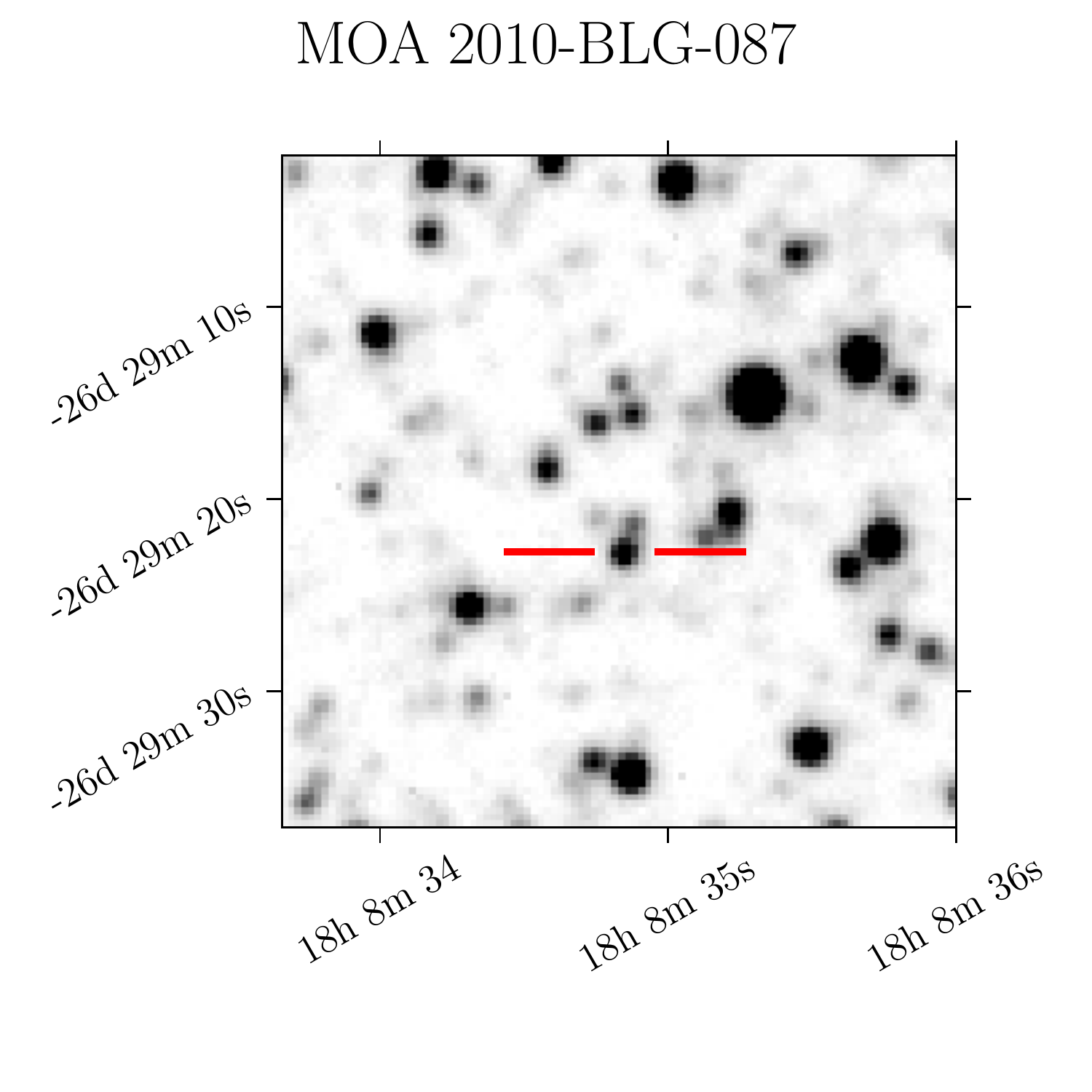}
\caption{Finder chart for MOA\,2010-BLG-087, created from one hour of ULTRACAM $g'$-band data on April 21. The seeing was 1". Note that the target was still in outburst and so appears brighter than its usual level.
}
\label{fig:moa-finder}
\end{center}
\end{figure}

\begin{figure}
\begin{center}
\includegraphics[width=\columnwidth]{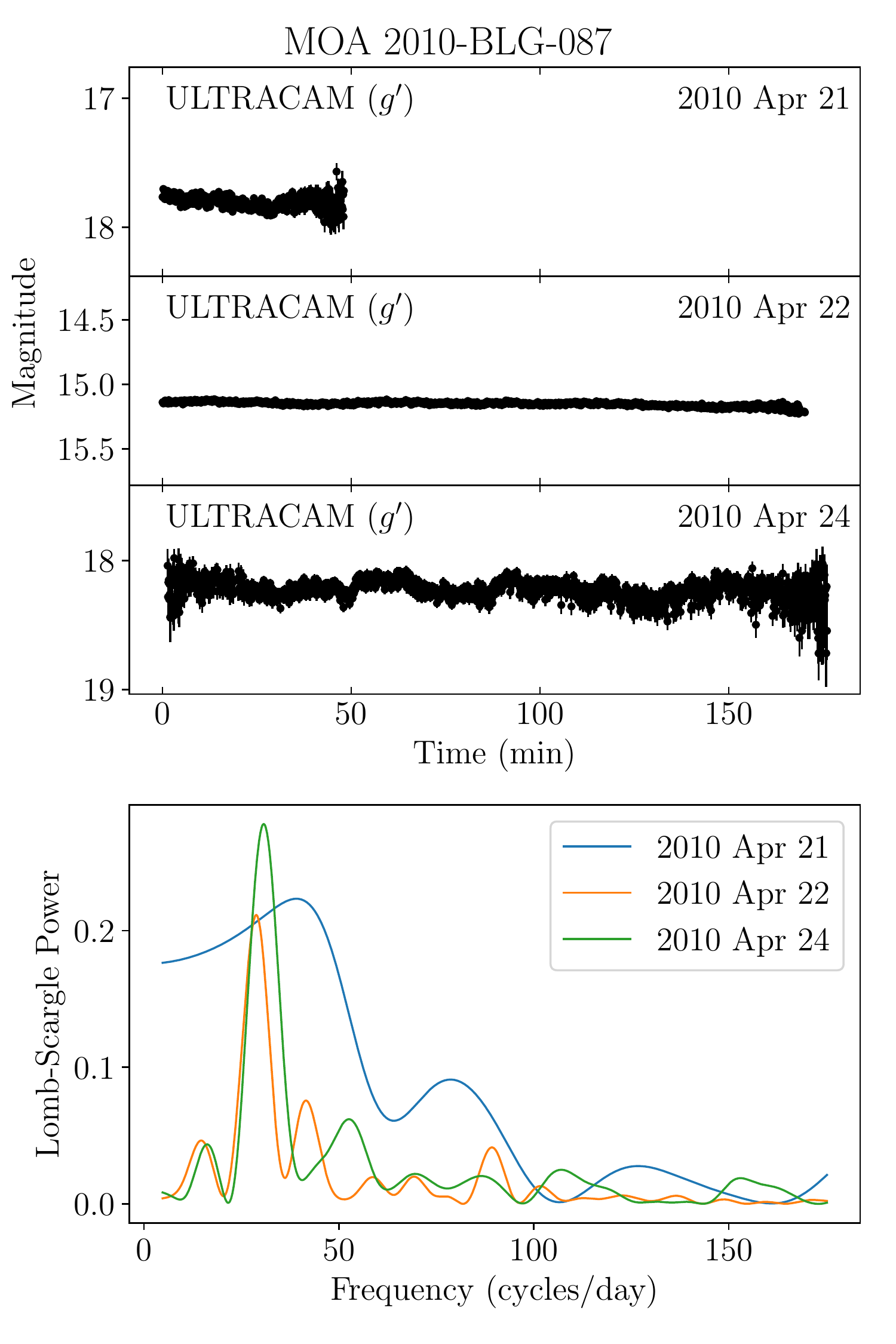}
\caption{\textit{Upper panels:} ULTRACAM photometry of MOA\,2010-BLG-087 from three nights. These observations were made shortly after outburst, and an echo outburst took place on April 22. \textit{Lower panel:} Lomb-Scargle periodograms of the data shown in the upper panels, after subtracting a flat baseline (April 21, due to its shorter coverage) or a three-term polynomial (April 22 and 24). The signal strengths between April 22 and 24 are comparable in terms of absolute flux, but the increased total brightness of the system on April 22 causes the signal to appear weaker on a magnitude scale.
}
\label{fig:moa-phot}
\end{center}
\end{figure}

We obtained photometry using ULTRACAM on three nights in 2010 April (Figure~\ref{fig:moa-phot}). The target was still at an elevated brightness, and the second night (April 22) coincided with an echo outburst that made the target three magnitudes brighter than April 21 and April 24.
Modulations with periods of 49.9\,min and 46.9\,min are visible on April 22 and 24 respectively. Given the short observation window on April 21, a similar signal would not be detected.

For two unresolved stars where one is variable, there will be a correlation between total brightness and the centroid position of the blended object.
We compared the \textit{x}- and \textit{y}-offsets in pixels of the target from the nearby comparison star (J2000 coordinates 18:08:35.3 $-$26:29:14.7) on the nights of April 22 and 24.
The \textit{y}-offset was particularly strongly correlated with \textit{g'}-band magnitude. 
Based on a Pearson's \textit{R} test, the two-tailed probabilities of such correlations occurring in an uncorrelated sample were $<10^{-26}$ and $<10^{-75}$, for April 22 and 24 respectively.
We also found a correlation between \textit{g'}-band magnitude and \textit{g'}-\textit{r'} colour, giving two-tailed probabilities of $<10^{-18}$ and $<10^{-44}$.
We therefore believe that the target AM\,CVn is blended with a nearby source.

In \textit{Gaia} Data Release 2 \citep{GaiaCollaboration2018}, two objects are resolved at these coordinates with a separation of 2" and magnitudes of $G=18.3$ and 19.4.
While it is tempting to suggest that these may be the two objects which are blended from the ground, we do not believe this to be the case, as a separation of 2" should be easily resolved by ULTRACAM.
We note that a nearby star is visible approximately 2" north of the target in Figure~\ref{fig:moa-finder}.

MOA\,2010-BLG-087 is the second AM\,CVn binary to have a co-aligned, late-type, main sequence star, after V407\,Vul. 
For V407\,Vul it is not known whether the co-aligned G-type star is a companion in a tertiary system or is aligned by coincidence, as the sky separation appears very small \citep[0.027";][]{Barros2007}.



\subsection{CRTS\,J1028$-$0819}

\begin{figure}
\begin{center}
\includegraphics[width=\columnwidth]{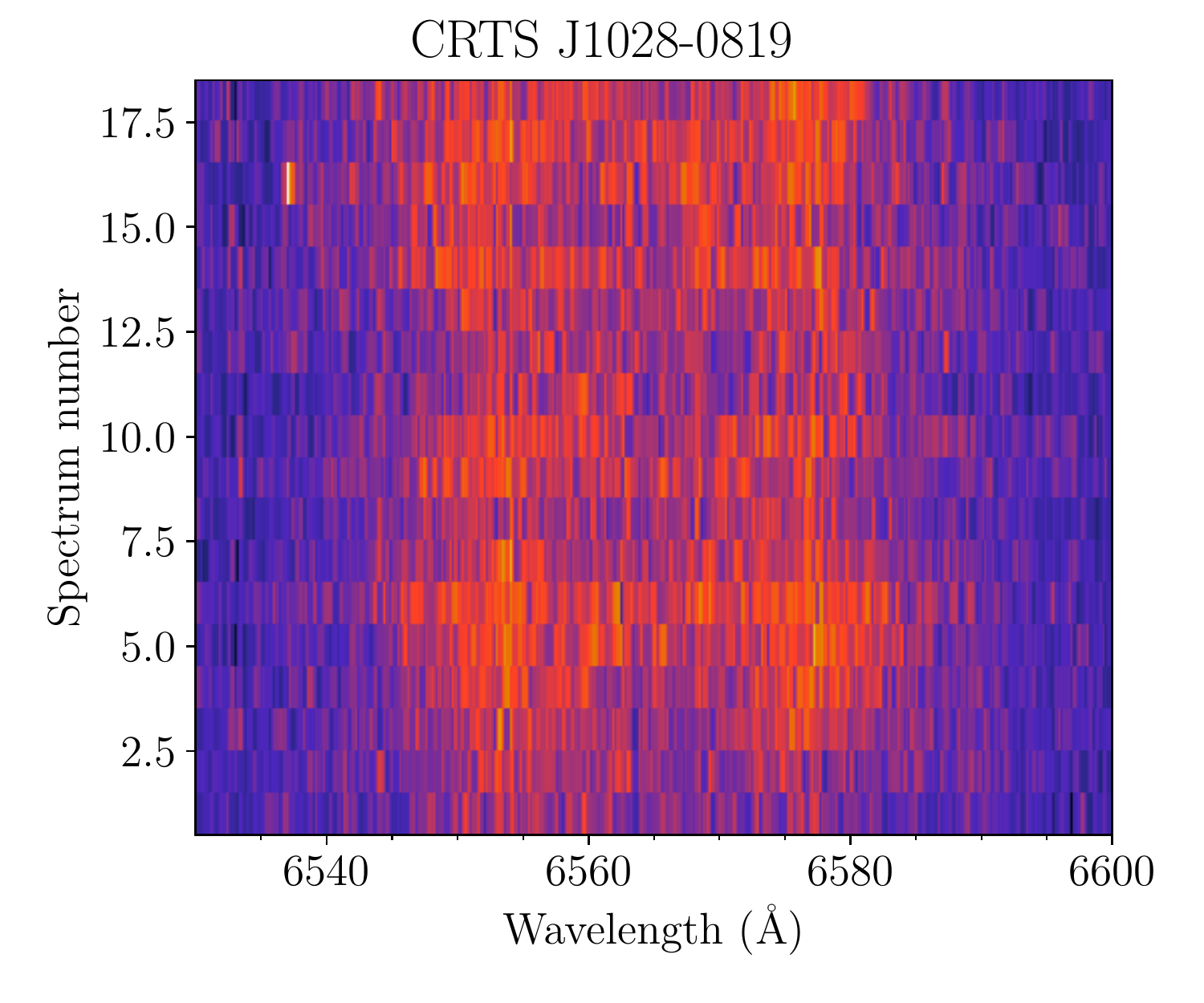}
\caption{Trailed spectra of the \halpha\ emission line of CRTS\,J1028$-$0819, spanning 120\,min. Each spectrum has been continuum-subtracted.
}
\label{fig:j1028-trail}
\end{center}
\end{figure}

\begin{figure}
\begin{center}
\includegraphics[width=\columnwidth]{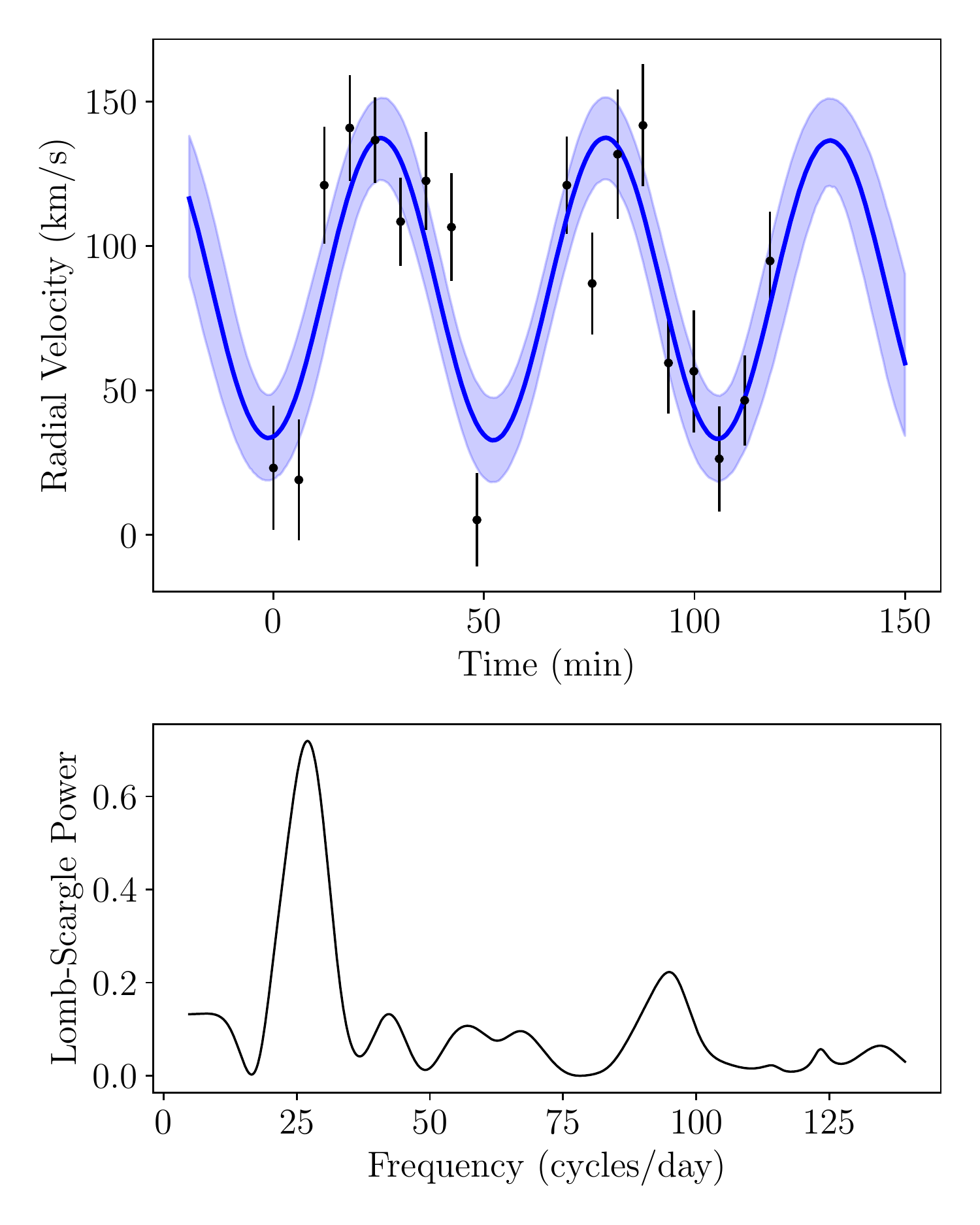}
\caption{\textit{Above:} RV measurements of the \halpha\ line in CRTS\,J1028$-$0819, measured by cross-correlating a Gaussian function with the emission lines. A sinusoidal fit is shown, with a period of $53.4 \pm 1.3$\,min. The shaded region shows the 2-$\sigma$ range of expected values produced by the MCMC fit.
\textit{Below:} A Lomb-Scargle plot produced from these RV measurements.}
\label{fig:j1028-sine}
\end{center}
\end{figure}

CRTS\,J1028$-$0819 (also known as CSS\,090331) was discovered in outburst by CRTS in 2009. A Stage B superhump period of $54.9 \pm 0.2$\,min was measured during outburst \citep{Kato2009}. \citet{Woudt2012} measured a photometric period during quiescence of $52.1 \pm 0.6$\,min, which they suggest to be the orbital period.
These photometric measurements suggest that CRTS\,J1028$-$0819 is the He CV with the shortest orbital period.

\subsubsection{Long-term Lightcurve}

The long-term lightcure of CRTS\,J1028$-$0819 (Figure~\ref{fig:longterm-lc}) shows a series of outbursts with an approximate recurrence time of 210 days.
An AM\,CVn binary at this orbital period would be expected to have a very low accretion rate. 
As such the presence of any outbursts would be remarkable, and if they did occur the expected recurrence time would be significantly longer \citep{Levitan2015}.
In terms of outbursts, it seems this He CV acts very unlike an AM\,CVn binary.

\subsubsection{Spectroscopy}

The mean spectrum of CRTS\,J1028$-$0819 is shown in Figure~\ref{fig:avspec-xshooter}. It shows a series of hydrogen and helium emission lines, as well as emission from calcium, sodium, magnesium, and nitrogen. 
CRTS\,J1028$-$0819 shows \heii\ emission at 4686\,\AA.

Trailed spectra of the \halpha\ line in CRTS\,J1028$-$0819 are shown in Figure~\ref{fig:j1028-trail}. The system does not appear to show any bright spot S-wave, and no central spike is visible. 
In order to search for RV variations, the \halpha\ line in each spectrum was cross-correlated with a Gaussian function, fixed such that its full-width at half-maximum was 1600\,km/s.
The resulting RV measurements are shown in Figure~\ref{fig:j1028-sine}.
A sinusoidal MCMC fit to these measurements produces a period of $53.4 \pm 1.3$\,min, consistent with the photometric periods measured by \citet{Kato2009} and \citet{Woudt2012}.

If we follow the interpretations of \citet{Kato2009} and \citet{Woudt2012} and take 52.1\,min as the orbital period and 54.9\,min as the stage B superhump period, Equation~\ref{eq:mcallisterB} gives $q = 0.25 \pm 0.06$, which falls within the range of $q$ measurements for other He~CVs.



\subsection{V418\,Ser}

V418\,Ser was classed as a dwarf nova following its outburst in 2004 \citep{Rykoff2004} but few follow-up observations were performed at the time. Following a CRTS-detected outburst in 2014, photometry showed the system to have a superhump period of $\approx 64.3$\,min \citep{Kato2015}. A spectrum reported by \citet{Garnavich2014} detected the presence of hydrogen.

\subsubsection{Long-term Lightcurve}

The long-term lightcurve of V418\,Ser (Figure~\ref{fig:longterm-lc}) shows a series of outbursts. The recurrence time is approximately 360 days, with a reasonable amount of scatter.
As was the case for CRTS\,J1028$-$0819, the presence of outbursts marks a clear departure from the behaviour that would be expected of an AM\,CVn binary at this orbital period.

\subsubsection{Spectroscopy}

The mean FORS2 spectrum is shown in Figure~\ref{fig:avspec-fors2}. The system shows a series of emission lines from hydrogen and helium. Emission is also seen from magnesium. 
The \hbeta\ line shows a notable absorption wing on its blue-ward side (the red-ward wing is hidden by nearby \hei\ emission).
The breadth of this absorption suggests it originates from the white dwarf photosphere.
No absorption wings are visible around the helium lines.

Trailed spectra of V418\,Ser show no visible S-wave from any bright spot or central spike, and no RV excursions were measurable for any emission line.
The lack of any visible velocity excursions in V418\,Ser suggests a low-mass donor star. 
This would imply a low mass-transfer rate, perhaps explaining the lack of any detectable bright spot in the spectroscopy.
The visibility of photospheric hydrogen absorption may imply that the disc is unusually faint relative to the accreting white dwarf, which would also be consistent with a low accretion rate into the disc. 

\subsubsection{High-Speed Photometry}

\begin{figure}
\begin{center}
\includegraphics[width=\columnwidth]{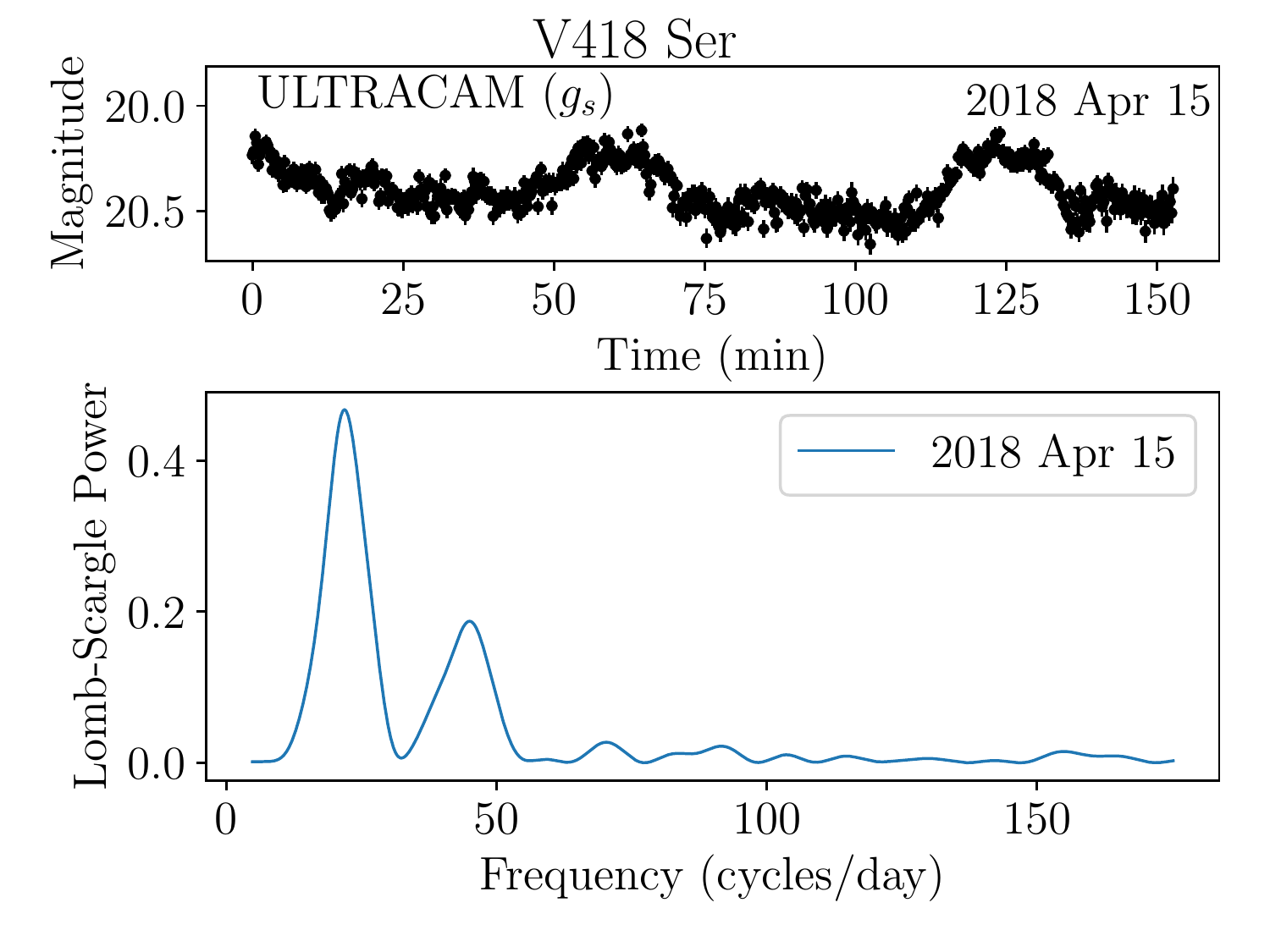}
\caption{\textit{Above:} ULTRACAM photometry of V418\,Ser. \textit{Below:} Lomb-Scargle periodogram of the data in the upper panel.
}
\label{fig:v418-phot}
\end{center}
\end{figure}

We obtained 2.5\,hours of photometry of V418\,Ser using ULTRACAM on 2018 April 15 (Figure~\ref{fig:v418-phot}). 
The lightcurve shows a feature resembling an orbital `hump' (centred on 60\,min and again on 125\,min in Figure~\ref{fig:v418-phot}). Similar features are seen in a number of CVs and result from the non-isotropic nature of emission from the bright spot \citep[eg.][]{Wood1986a}
The presence of such a signal generally implies a relatively edge-on orbital inclination.
However, if this signal does result from the bright spot, it is difficult to reconcile its strength with the weakness of the bright spot in spectroscopy.

The period of this modulation is $65.9 \pm 0.6$\,min, where the uncertainty was estimated by fitting a sinusoid to the data.
This is similar to the superhump period observed by \citet{Kato2015}, but slightly longer.
As no uncertainty was reported on the superhump period, we are unable to say if the two periods are consistent.
If not, this may be an example of a system with negative superhumps -- a superhump signal for which the period is slightly shorter than the orbital period rather than longer.
Negative superhumps originate from nodal precession of an eccentric disc (as opposed to apsidal precession in the positive superhump case) and are observed in a minority of superhumping systems \citep{Hellier2001}.


\section{Discussion}
\label{sec:6-discussion}

\begin{table}
\caption{Orbital periods, superhump periods, and mass ratios for the target systems. Periods are in minutes. 
Periods marked $\dagger$ are previously published (see the references in relevant sections of the text).
For CRTS\,J1028$-$0819, we show both the spectroscopic orbital period and the suggested photometric periods.
\cmnt{Check all these periods again before publishing!?}
}
\label{tab:6-results}
\begin{tabular}{lccccc}
Target & $P_\mathrm{orb}$& $P_\mathrm{sh}$ & $q$\\
\hline
SDSS\,J1505		& $67.8 \pm 2.2$ 	& -- 					& $0.011 ^{+ 0.003} _{- 0.001}$\\

ASASSN-14ei				& 41--42.5  	& -- 		& -- \\

ASASSN-14mv				& -- 				& $40.9 \pm 0.3$ 		& -- \\

MOA\,2010		& --				& $\approx 47$--50			& -- \\

CRTS\,J1028		& $53.4 \pm 1.3$ 	\\
\hspace{2em} (phot)				& $52.1 \pm 0.6 ^ \dagger$ & $54.9 \pm 0.2 ^ \dagger$		& $0.25 \pm 0.06$ \\

V418\,Ser				& $65.9 \pm 0.6$	& $\approx 64.3 ^ \dagger$ 			& --\\
\hline
\end{tabular}
\end{table}

\begin{figure}
\begin{center}
\includegraphics[width=\columnwidth]{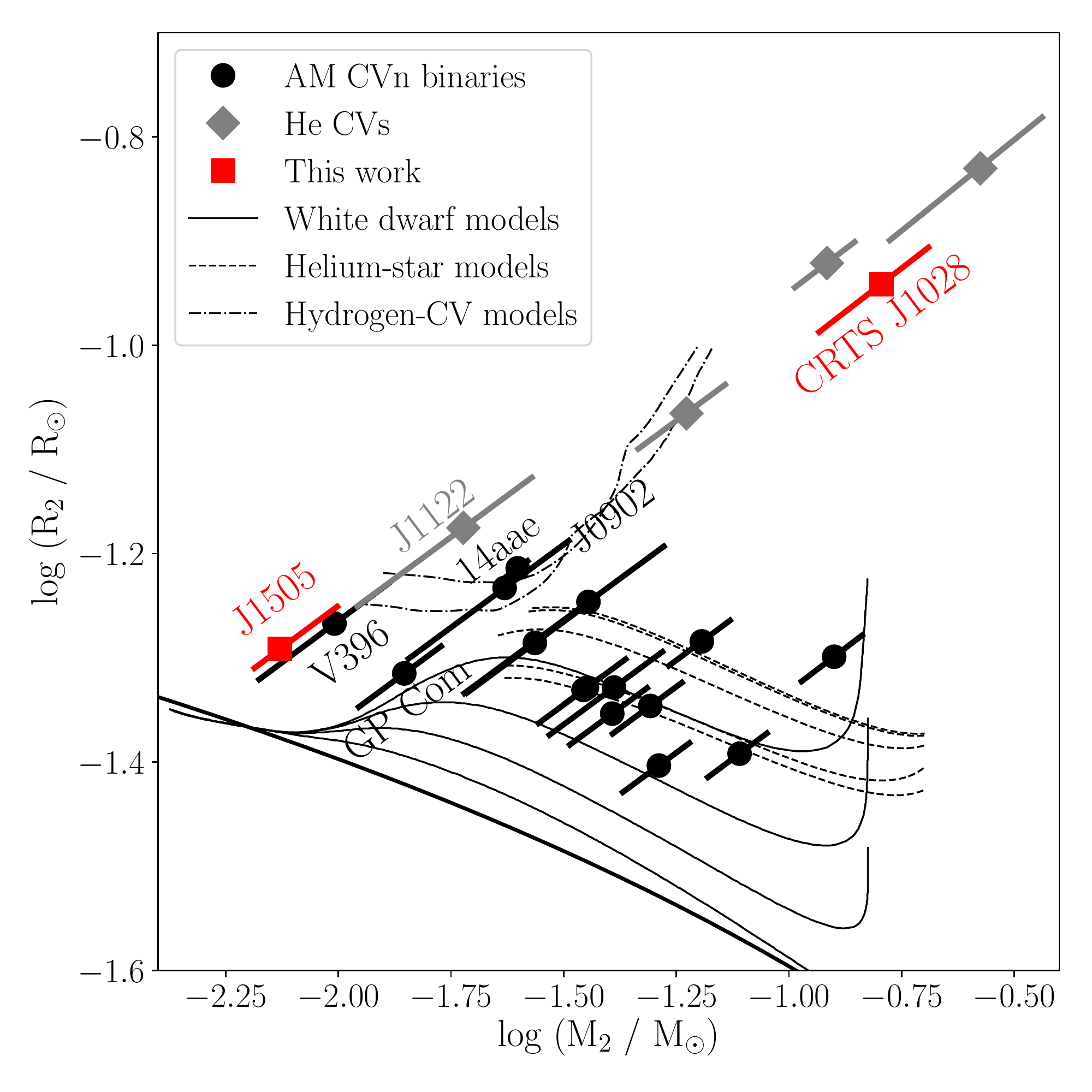}
\caption{The positions of known AM\,CVn binaries and He~CVs in donor mass-radius space. 
AM\,CVn data are taken from \citep[and references therein]{Green2018b}, and He~CV data are from the references in Table~\ref{tab:hecvs}.
The $q$ measurements made in this work are highlighted in red and labelled.
Model evolutionary tracks shown are taken from \citep[white dwarf donors]{Deloye2007}, \citep[helium star donors]{Yungelson2008}, and \citep[evolved CVs]{Goliasch2015}.
}
\label{fig:mass-radius-6}
\end{center}
\end{figure}

\subsection{Photometric Behaviour}

As discussed in Section~\ref{sec:14ei}, the photometric behaviour of ASASSN-14ei is remarkable.
The increased brightness and increased rate of outbursts following the 2014 superoutburst of ASASSN-14ei strongly suggest that the mass transfer rate increased in response to the 2014 superoutburst, perhaps due to heating of the donor star.
The system seems even more remarkable when compared to ASASSN-14mv. The two systems appear to be similar in most measurable ways (the orbital periods and outburst magnitudes are similar, and both show the same echo outburst behaviour). 
Why then does ASASSN-14ei show an increase in mass transfer rate while ASASSN-14mv does not?
This difference in behaviour may hint at some underlying difference in the nature of the two binaries, perhaps in the nature of the donor stars.

Leaving ASASSN-14ei aside, there is a stark difference between the outburst behaviour of the AM\,CVn binaries in this sample (which undergo outbursts rarely if at all) and that of the He~CVs (which undergo outbursts with recurrence timescales on the order of 100s of days).
The more frequent outbursts of the He~CVs are likely due to higher accretion rates, resulting from higher mass donors.

\subsection{Evolutionary Natures}

One of the motivations behind this paper was to examine the possibility of an evolutionary link between AM\,CVn binaries and He~CVs.
A key signifier of the evolutionary history of these binaries is the donor mass and radius.
In Figure~\ref{fig:mass-radius-6} we show the donor masses and radii for all AM\,CVn binaries and He~CVs with appropriate measurements. 
AM\,CVn measurements are taken from \citet[their Table~4 and references therein]{Green2018b}, and He~CV measurements from the sources listed in Table~\ref{tab:hecvs}.
For the majority of these binaries only the mass ratio is known. For these binaries we assume $M_1 = 0.7 \pm 0.1 M_\odot$.
Uncertainties are plotted diagonally due to the tight correlation between donor mass and radius, originating from the relationship between orbital period and donor mean density \citep{Faulkner1972a}.
By this relationship, binaries with the shortest orbital periods will appear in the bottom right of the figure, while the longest orbital periods would appear in the top left.
On the same figure, we plot mass-radius evolutionary tracks for AM\,CVn binaries descended from the white dwarf donor channel \citep{Deloye2007}, the helium star donor channel \citep{Yungelson2008}, and the evolved CV channel \citep{Goliasch2015}. 

The two populations follow different distributions. 
He~CVs are constrained to the 50-70\,min orbital period range, but their distribution extends to much higher donor masses than AM\,CVn binaries of the same orbital period. 
There is, however, a region of mass-radius space in which low-mass He~CV donors overlap with long-period AM\,CVn donors.
The He~CV CRTS\,J1122$-$1110 and the AM\,CVn binaries Gaia14aae, SDSS\,J0902+3819, and V396\,Hya are all close to this region of overlap.

The $q$ measurements made in this paper are highlighted in red in Figure~\ref{fig:mass-radius-6}.
CRTS\,J1028$-$0819 appears to be towards the high-donor mass end of the distribution of He~CVs, as would be expected based on its relatively high outburst rate.
SDSS\,J1505+0659 appears to have a very similar donor to the next-longest-period AM\,CVn binary, V396\,Hya. 
Looking at the five longest-period AM\,CVn binaries (five left-most in Figure~\ref{fig:mass-radius-6}), one could suggest a bifurcation in their radius distribution. SDSS\,J1505+0659, V396\,Hya and GP\,Com all appear to have significantly smaller radii for their orbital period than Gaia14aae and SDSS\,J0902+3819, which are more inflated.
Such a bifurcation is merely suggestive and is not detected to any significance given the uncertainties on these measurements.

\section{Conclusions}

We have presented a spectroscopic and photometric study of six accreting binary systems with orbital periods shorter than the period minimum for hydrogen-accreting CVs.
Four of these binaries are of the AM\,CVn type, and two are He~CVs.
Four of these binaries have measured orbital periods, and four have measured superhump periods, either resulting from our observations or previously published.
The binaries in this sample include the AM\,CVn binary with the longest known orbital period (SDSS\,J1505+0659 at $67.8 \pm 2.2$\,min) and the He~CV with the shortest known orbital period (CRTS\,J1028$-$0819 at $53.4 \pm 1.3$\,min).

All but SDSS\,J1505+0659 have been detected in outburst. 
The outburst rates of the two He~CVs are significantly higher than the outburst rates of the AM\,CVn binaries, likely due to higher-mass donor stars.
Two of the AM\,CVn binaries, ASASSN-14ei and ASASSN-14mv, have notable outburst behaviour, with 12 and 10 echo outbursts respectively.
The mass transfer rate of ASASSN-14ei appears to have increased substantially during the 2014 outburst, causing an increase in brightness and outburst rate. The system has since been gradually returning towards its quiescent state.

We measured the mass ratio for SDSS\,J1505+0659, which has a low-mass donor similar to that of V396\,Vul.
We also estimate the mass ratio of CRTS\,J1028$-$0819 based on previously-published photometric periods, and find that it has a high-mass donor, consistent with other He~CV measurements.
We note that there is some overlap between the regions of donor mass-radius space populated by AM\,CVn binaries and He~CVs (Figure~\ref{fig:mass-radius-6}), but the systems studied in this work are not in that region of overlap.

\section*{Acknowledgements}

\cmnt{Thank the reviewer?}
The authors would like to thank the anonymous reviewer for their feedback. We would also like to thank Christian Knigge and Pier-Emmanuel Tremblay for helpful discussions and comments (during a viva examination!).
We acknowledge with thanks the variable star observations from the AAVSO International Database contributed by observers worldwide and used in this research, including Franz-Josef Hambsch, Gordon Myers, James Boardman, and David Cejudo Fernandez.

MJG acknowledges funding from STFC via studentship grant ST/N504506/1. 
MJG, TRM and DS are funded by STFC grant ST/P000495/1. 
VSD, SPL, ULTRACAM, and ULTRASPEC are funded by STFC via consolidated grant ST/J001589.
SGP acknowledges support from the STFC via an Ernest Rutherford Fellowship. 

This work is based in part on observations collected at the European Organisation for Astronomical Research in the Southern Hemisphere under ESO programmes 085.D-0541, 095.D-0888, 0100.D-0425, 0101.D-0852, 0102.D-0060, and 0102.D-0953. 
Data were also obtained using the 2.4\,m Thai National Telescope (TNT) operated by the National Astronomy Research Institute of Thailand (NARIT). 
Further data were obtained under proposal ITP8 using the 4.2\,m William Herschel Telescope (WHT), operated by the Isaac Newton Group of Telescopes. 

This work made use of software packages \program{scipy}, \program{numpy}, \program{astropy}, \program{matplotlib}, \program{emcee}, \program{pamela}, and \program{molly}.

\bibliographystyle{mnras}
\bibliography{refs} 

\begin{thebibliography}{}
\makeatletter
\relax
\def\mn@urlcharsother{\let\do\@makeother \do\$\do\&\do\#\do\^\do\_\do\%\do\~}
\def\mn@doi{\begingroup\mn@urlcharsother \@ifnextchar [ {\mn@doi@}
  {\mn@doi@[]}}
\def\mn@doi@[#1]#2{\def\@tempa{#1}\ifx\@tempa\@empty \href
  {http://dx.doi.org/#2} {doi:#2}\else \href {http://dx.doi.org/#2} {#1}\fi
  \endgroup}
\def\mn@eprint#1#2{\mn@eprint@#1:#2::\@nil}
\def\mn@eprint@arXiv#1{\href {http://arxiv.org/abs/#1} {{\tt arXiv:#1}}}
\def\mn@eprint@dblp#1{\href {http://dblp.uni-trier.de/rec/bibtex/#1.xml}
  {dblp:#1}}
\def\mn@eprint@#1:#2:#3:#4\@nil{\def\@tempa {#1}\def\@tempb {#2}\def\@tempc
  {#3}\ifx \@tempc \@empty \let \@tempc \@tempb \let \@tempb \@tempa \fi \ifx
  \@tempb \@empty \def\@tempb {arXiv}\fi \@ifundefined
  {mn@eprint@\@tempb}{\@tempb:\@tempc}{\expandafter \expandafter \csname
  mn@eprint@\@tempb\endcsname \expandafter{\@tempc}}}

\bibitem[\protect\citeauthoryear{Abazajian et~al.,}{Abazajian
  et~al.}{2009}]{Abazajian2009}
Abazajian K.~N.,  et~al., 2009, \mn@doi [The Astrophysical Journal Supplement
  Series] {10.1088/0067-0049/182/2/543}, 182, 543

\bibitem[\protect\citeauthoryear{Appenzeller et~al.,}{Appenzeller
  et~al.}{1998}]{Appenzeller1998}
Appenzeller I.,  et~al., 1998, The Messenger, 94, 1

\bibitem[\protect\citeauthoryear{Armstrong, Patterson  \& Kemp}{Armstrong
  et~al.}{2012}]{Armstrong2012}
Armstrong E.,  Patterson J.,   Kemp J.,  2012, \mn@doi [Monthly Notices of the
  Royal Astronomical Society] {10.1111/j.1365-2966.2012.20463.x}, 421, 2310

\bibitem[\protect\citeauthoryear{Augusteijn, van Kerkwijk  \& van
  Paradijs}{Augusteijn et~al.}{1993}]{Augusteijn1993}
Augusteijn T.,  van Kerkwijk M.~H.,   van Paradijs J.,  1993, Astronomy and
  Astrophysics, 267, L55

\bibitem[\protect\citeauthoryear{Augusteijn, van~der Hooft, de Jong  \& van
  Paradijs}{Augusteijn et~al.}{1996}]{Augusteijn1996}
Augusteijn T.,  van~der Hooft F.,  de Jong J.~A.,   van Paradijs J.,  1996,
  Astronomy and Astrophysics, 311, 889

\bibitem[\protect\citeauthoryear{{Barros} et~al.,}{{Barros}
  et~al.}{2007}]{Barros2007}
{Barros} S.~C.~C.,  et~al., 2007, \mn@doi [Monthly Notices of the Royal
  Astronomical Society] {10.1111/j.1365-2966.2006.11244.x}, \href
  {https://ui.adsabs.harvard.edu/abs/2007MNRAS.374.1334B} {374, 1334}

\bibitem[\protect\citeauthoryear{Breedt}{Breedt}{2015}]{Breedt2015}
Breedt E.,  2015, \mn@doi [Proceedings of The Golden Age of Cataclysmic
  Variables and Related Objects - III (Golden2015)] {2015gacv.workE..25B},
  p.~25

\bibitem[\protect\citeauthoryear{Breedt, G{\"{a}}nsicke, Marsh, Steeghs, Drake
  \& Copperwheat}{Breedt et~al.}{2012}]{Breedt2012}
Breedt E.,  G{\"{a}}nsicke B.~T.,  Marsh T.~R.,  Steeghs D.,  Drake A.~J.,
  Copperwheat C.~M.,  2012, \mn@doi [Monthly Notices of the Royal Astronomical
  Society] {10.1111/j.1365-2966.2012.21724.x}, 425, 2548

\bibitem[\protect\citeauthoryear{Breedt et~al.,}{Breedt
  et~al.}{2014}]{Breedt2014}
Breedt E.,  et~al., 2014, \mn@doi [Monthly Notices of the Royal Astronomical
  Society] {10.1093/mnras/stu1377}, 443, 3174

\bibitem[\protect\citeauthoryear{Breivik, Kremer, Bueno, Larson, Coughlin  \&
  Kalogera}{Breivik et~al.}{2018}]{Breivik2018}
Breivik K.,  Kremer K.,  Bueno M.,  Larson S.~L.,  Coughlin S.,   Kalogera V.,
  2018, \mn@doi [The Astrophysical Journal Letters] {10.3847/2041-8213/aaaa23},
  854, 1

\bibitem[\protect\citeauthoryear{Cannizzo \& Nelemans}{Cannizzo \&
  Nelemans}{2015}]{Cannizzo2015}
Cannizzo J.~K.,  Nelemans G.,  2015, \mn@doi [The Astrophysical Journal]
  {10.1088/0004-637X/803/1/19}, 803, 19

\bibitem[\protect\citeauthoryear{Cannizzo \& Ramsay}{Cannizzo \&
  Ramsay}{2019}]{Cannizzo2019}
Cannizzo J.~K.,  Ramsay G.,  2019, \mn@doi [The Astronomical Journal]
  {10.3847/1538-3881/ab04ac}, 157, 130

\bibitem[\protect\citeauthoryear{Carter et~al.,}{Carter
  et~al.}{2013}]{Carter2013}
Carter P.~J.,  et~al., 2013, \mn@doi [Monthly Notices of the Royal Astronomical
  Society] {10.1093/mnras/sts485}, 429, 2143

\bibitem[\protect\citeauthoryear{Carter et~al.,}{Carter
  et~al.}{2014}]{Carter2014}
Carter P.~J.,  et~al., 2014, \mn@doi [Monthly Notices of the Royal Astronomical
  Society] {10.1093/mnras/stu142}, 439, 2848

\bibitem[\protect\citeauthoryear{Chambers et~al.,}{Chambers
  et~al.}{2016}]{Chambers2016}
Chambers K.~C.,  et~al., 2016, preprint (\mn@eprint {arXiv} {1612.05560})

\bibitem[\protect\citeauthoryear{Chochol et~al.,}{Chochol
  et~al.}{2015}]{Chochol2015}
Chochol D.,  et~al., 2015, \mn@doi [Acta Polytechnica CTU Proceedings]
  {10.14311/APP.2015.02.0165}, 2, 165

\bibitem[\protect\citeauthoryear{Copperwheat et~al.,}{Copperwheat
  et~al.}{2011}]{Copperwheat2011}
Copperwheat C.~M.,  et~al., 2011, \mn@doi [Monthly Notices of the Royal
  Astronomical Society] {10.1111/j.1365-2966.2010.17508.x}, 410, 1113

\bibitem[\protect\citeauthoryear{Deloye, Taam, Winisdoerffer  \&
  Chabrier}{Deloye et~al.}{2007}]{Deloye2007}
Deloye C.~J.,  Taam R.~E.,  Winisdoerffer C.,   Chabrier G.,  2007, \mn@doi
  [Monthly Notices of the Royal Astronomical Society]
  {10.1111/j.1365-2966.2007.12262.x}, 381, 525

\bibitem[\protect\citeauthoryear{Denisenko et~al.,}{Denisenko
  et~al.}{2014}]{Denisenko2014}
Denisenko D.,  et~al., 2014, The Astronomer's Telegram, 6857, 1

\bibitem[\protect\citeauthoryear{{Denisenkov} \& {Ivanov}}{{Denisenkov} \&
  {Ivanov}}{1987}]{Denisenkov1987}
{Denisenkov} P.~A.,  {Ivanov} V.~V.,  1987, Soviet Astronomy Letters, \href
  {https://ui.adsabs.harvard.edu/abs/1987SvAL...13..214D} {13, 214}

\bibitem[\protect\citeauthoryear{Dhillon et~al.,}{Dhillon
  et~al.}{2007}]{ULTRACAM}
Dhillon V.~S.,  et~al., 2007, \mn@doi [Monthly Notices of the Royal
  Astronomical Society] {10.1111/j.1365-2966.2007.11881.x}, 378, 825

\bibitem[\protect\citeauthoryear{Dhillon et~al.,}{Dhillon
  et~al.}{2014}]{ULTRASPEC}
Dhillon V.~S.,  et~al., 2014, \mn@doi [Monthly Notices of the Royal
  Astronomical Society] {10.1093/mnras/stu1660}, 444, 4009

\bibitem[\protect\citeauthoryear{{Dhillon} et~al.,}{{Dhillon}
  et~al.}{2018}]{Dhillon2018}
{Dhillon} V.,  et~al., 2018, in \procspie. p. 107020L (\mn@eprint {arXiv}
  {1807.00557}), \mn@doi{10.1117/12.2312041}

\bibitem[\protect\citeauthoryear{Drake et~al.,}{Drake et~al.}{2009}]{Drake2009}
Drake A.~J.,  et~al., 2009, \mn@doi [Astrophysical Journal]
  {10.1088/0004-637X/696/1/870}, 696, 870

\bibitem[\protect\citeauthoryear{Eggleton}{Eggleton}{1983}]{Eggleton1983}
Eggleton P.~P.,  1983, \mn@doi [The Astrophysical Journal] {10.1086/160960},
  268, 368

\bibitem[\protect\citeauthoryear{Faulkner, Flannery  \& Warner}{Faulkner
  et~al.}{1972}]{Faulkner1972a}
Faulkner J.,  Flannery B.~P.,   Warner B.,  1972, \mn@doi [The Astrophysical
  Journal] {10.1086/180989}, 175, L79

\bibitem[\protect\citeauthoryear{Fontaine et~al.,}{Fontaine
  et~al.}{2011}]{Fontaine2011}
Fontaine G.,  et~al., 2011, \mn@doi [The Astrophysical Journal]
  {10.1088/0004-637X/726/2/92}, 726, 92

\bibitem[\protect\citeauthoryear{Foreman-Mackey, Hogg, Lang  \&
  Goodman}{Foreman-Mackey et~al.}{2013}]{Foreman-Mackey2013}
Foreman-Mackey D.,  Hogg D.~W.,  Lang D.,   Goodman J.,  2013, \mn@doi
  [Publications of the Astronomical Society of Pacific] {10.1086/670067}, 125,
  306

\bibitem[\protect\citeauthoryear{{Gaia Collaboration} et~al.,}{{Gaia
  Collaboration} et~al.}{2018}]{GaiaCollaboration2018}
{Gaia Collaboration} T.,  et~al., 2018, \mn@doi [Astronomy & Astrophysics]
  {10.1051/0004-6361/201833051}, 616, A1

\bibitem[\protect\citeauthoryear{Garnavich, Littlefield, Terndrup  \&
  Adams}{Garnavich et~al.}{2014}]{Garnavich2014}
Garnavich P.,  Littlefield C.,  Terndrup D.,   Adams S.,  2014, The
  Astronomer's Telegram, 6287

\bibitem[\protect\citeauthoryear{Goliasch \& Nelson}{Goliasch \&
  Nelson}{2015}]{Goliasch2015}
Goliasch J.,  Nelson L.,  2015, \mn@doi [The Astrophysical Journal]
  {10.1088/0004-637X/809/1/80}, 809, 80

\bibitem[\protect\citeauthoryear{Goodman \& Weare}{Goodman \&
  Weare}{2010}]{Goodman2010}
Goodman J.,  Weare J.,  2010, \mn@doi [Communications in Applied Mathematics
  and Computational Science] {10.2140/camcos.2010.5.65}, 5, 65

\bibitem[\protect\citeauthoryear{Green et~al.,}{Green
  et~al.}{2018a}]{Green2018}
Green M.~J.,  et~al., 2018a, \mn@doi [Monthly Notices of the Royal Astronomical
  Society] {10.1093/mnras/sty299}, 476, 1663

\bibitem[\protect\citeauthoryear{Green et~al.,}{Green
  et~al.}{2018b}]{Green2018b}
Green M.~J.,  et~al., 2018b, \mn@doi [Monthly Notices of the Royal Astronomical
  Society] {10.1093/mnras/sty1032}, 477, 5646

\bibitem[\protect\citeauthoryear{Green et~al.,}{Green et~al.}{2019}]{Green2019}
Green M.~J.,  et~al., 2019, \mn@doi [Monthly Notices of the Royal Astronomical
  Society] {10.1093/mnras/stz469}, 485, 1947

\bibitem[\protect\citeauthoryear{Hardy et~al.,}{Hardy et~al.}{2017}]{Hardy2017}
Hardy L.~K.,  et~al., 2017, \mn@doi [Monthly Notices of the Royal Astronomical
  Society] {10.1093/mnras/stw3051}, 465, 4968

\bibitem[\protect\citeauthoryear{Hellier}{Hellier}{2001}]{Hellier2001}
Hellier C.,  2001, {Cataclysmic Variable Stars}.
Springer, Berlin

\bibitem[\protect\citeauthoryear{Iben \& Tutukov}{Iben \&
  Tutukov}{1987}]{Iben1987}
Iben I.~J.,  Tutukov A.~V.,  1987, \mn@doi [The Astrophysical Journal]
  {10.1086/165011}, 313, 727

\bibitem[\protect\citeauthoryear{Imada, Isogai, Yanagisawa  \& Kawai}{Imada
  et~al.}{2018}]{Imada2018}
Imada A.,  Isogai K.,  Yanagisawa K.,   Kawai N.,  2018, \mn@doi [Publications
  of the Astronomical Society of Japan] {10.1093/pasj/psy074}, 70, 79

\bibitem[\protect\citeauthoryear{{Isogai}, {Kato}, {Monard}, {Hambsch},
  {Myers}, {Starr}, {Cook}  \& {Nogami}}{{Isogai} et~al.}{2019}]{Isogai2019}
{Isogai} K.,  {Kato} T.,  {Monard} B.,  {Hambsch} F.-J.,  {Myers} G.,  {Starr}
  P.,  {Cook} L.~M.,   {Nogami} D.,  2019, \mn@doi [\pasj]
  {10.1093/pasj/psz018}, \href
  {https://ui.adsabs.harvard.edu/abs/2019PASJ...71...48I} {71, 48}

\bibitem[\protect\citeauthoryear{Israel, Panzera, Campana, Lazzati, Covino,
  Tagliaferri  \& Stella}{Israel et~al.}{1999}]{Israel1999}
Israel G.~L.,  Panzera M.~R.,  Campana S.,  Lazzati D.,  Covino S.,
  Tagliaferri G.,   Stella L.,  1999, Astronomy and Astrophysics, 349, L1

\bibitem[\protect\citeauthoryear{Ivanova et~al.,}{Ivanova
  et~al.}{2013}]{Ivanova2013}
Ivanova N.,  et~al., 2013, \mn@doi [Astronomy and Astrophysics Review]
  {10.1007/s00159-013-0059-2}, 21, 59

\bibitem[\protect\citeauthoryear{Kato \& Osaki}{Kato \& Osaki}{2013}]{Kato2013}
Kato T.,  Osaki Y.,  2013, \mn@doi [Publications of the Astronomical Society of
  Japan] {10.1093/pasj/65.6.115}, 65, 115

\bibitem[\protect\citeauthoryear{Kato et~al.,}{Kato et~al.}{2004}]{Kato2004}
Kato T.,  et~al., 2004, \mn@doi [Publications of the Astronomical Society of
  Japan] {10.1093/pasj/56.sp1.S1}, 56, S1

\bibitem[\protect\citeauthoryear{Kato et~al.,}{Kato et~al.}{2009}]{Kato2009}
Kato T.,  et~al., 2009, \mn@doi [Publications of the Astronomical Society of
  Japan] {10.1093/pasj/61.sp2.S395}, 61, S395

\bibitem[\protect\citeauthoryear{Kato, Hambsch  \& Monard}{Kato
  et~al.}{2015a}]{Kato2015}
Kato T.,  Hambsch F.-J.,   Monard B.,  2015a, \mn@doi [Publications of the
  Astronomical Society of Japan] {10.1093/pasj/psv010}, 67, L2

\bibitem[\protect\citeauthoryear{Kato et~al.,}{Kato et~al.}{2015b}]{Kato2015a}
Kato T.,  et~al., 2015b, \mn@doi [Publications of the Astronomical Society of
  Japan] {10.1093/pasj/psv072}, 67, 105

\bibitem[\protect\citeauthoryear{Knigge}{Knigge}{2006}]{Knigge2006}
Knigge C.,  2006, \mn@doi [Monthly Notices of the Royal Astronomical Society]
  {10.1111/j.1365-2966.2006.11096.x}, 373, 484

\bibitem[\protect\citeauthoryear{Knigge, Baraffe  \& Patterson}{Knigge
  et~al.}{2011}]{Knigge2011}
Knigge C.,  Baraffe I.,   Patterson J.,  2011, \mn@doi [The Astrophysical
  Journal Supplement Series] {10.1088/0067-0049/194/2/28}, 194, 28

\bibitem[\protect\citeauthoryear{Kochanek et~al.,}{Kochanek
  et~al.}{2017}]{Kochanek2017}
Kochanek C.~S.,  et~al., 2017, \mn@doi [Publications of the Astronomical
  Society of the Pacific] {10.1088/1538-3873/aa80d9}, 129, 104502

\bibitem[\protect\citeauthoryear{Kotko, Lasota, Dubus  \& Hameury}{Kotko
  et~al.}{2012}]{Kotko2012}
Kotko I.,  Lasota J.-P.,  Dubus G.,   Hameury J.-M.,  2012, \mn@doi [Astronomy
  and Astrophysics] {10.1051/0004-6361/201219156}, 544, A13

\bibitem[\protect\citeauthoryear{Kremer, Breivik, Larson  \& Kalogera}{Kremer
  et~al.}{2017}]{Kremer2017}
Kremer K.,  Breivik K.,  Larson S.~L.,   Kalogera V.,  2017, \mn@doi [The
  Astrophysical Journal] {10.3847/1538-4357/aa8557}, 846, 95

\bibitem[\protect\citeauthoryear{Kupfer, Groot, Levitan, Steeghs, Marsh, Rutten
   \& Nelemans}{Kupfer et~al.}{2013}]{Kupfer2013}
Kupfer T.,  Groot P.~J.,  Levitan D.,  Steeghs D.,  Marsh T.~R.,  Rutten R.
  G.~M.,   Nelemans G.,  2013, \mn@doi [Monthly Notices of the Royal
  Astronomical Society] {10.1093/mnras/stt524}, 432, 2048

\bibitem[\protect\citeauthoryear{Kupfer et~al.,}{Kupfer
  et~al.}{2015}]{Kupfer2015}
Kupfer T.,  et~al., 2015, \mn@doi [Monthly Notices of the Royal Astronomical
  Society] {10.1093/mnras/stv1609}, 453, 483

\bibitem[\protect\citeauthoryear{Kupfer, Steeghs, Groot, Marsh, Nelemans  \&
  Roelofs}{Kupfer et~al.}{2016}]{Kupfer2016}
Kupfer T.,  Steeghs D.,  Groot P.~J.,  Marsh T.~R.,  Nelemans G.,   Roelofs G.
  H.~A.,  2016, \mn@doi [Monthly Notices of the Royal Astronomical Society]
  {10.1093/mnras/stw126}, 457, 1828

\bibitem[\protect\citeauthoryear{Kupfer et~al.,}{Kupfer
  et~al.}{2018}]{Kupfer2018}
Kupfer T.,  et~al., 2018, \mn@doi [Monthly Notices of the Royal Astronomical
  Society] {10.1093/mnras/sty1545}, 480, 302

\bibitem[\protect\citeauthoryear{Lasota}{Lasota}{2001}]{Lasota2001}
Lasota J.-P.,  2001, \mn@doi [New Astronomy Reviews]
  {10.1016/S1387-6473(01)00112-9}, 45, 449

\bibitem[\protect\citeauthoryear{Lawrence et~al.,}{Lawrence
  et~al.}{2007}]{Lawrence2007}
Lawrence A.,  et~al., 2007, \mn@doi [Monthly Notices of the Royal Astronomical
  Society] {10.1111/j.1365-2966.2007.12040.x}, 379, 1599

\bibitem[\protect\citeauthoryear{Levitan, Groot, Prince, Kulkarni, Laher, Ofek,
  Sesar  \& Surace}{Levitan et~al.}{2015}]{Levitan2015}
Levitan D.,  Groot P.~J.,  Prince T.~A.,  Kulkarni S.~R.,  Laher R.,  Ofek
  E.~O.,  Sesar B.,   Surace J.,  2015, \mn@doi [Monthly Notices of the Royal
  Astronomical Society] {10.1093/mnras/stu2105}, 446, 391

\bibitem[\protect\citeauthoryear{Littlefield et~al.,}{Littlefield
  et~al.}{2013}]{Littlefield2013}
Littlefield C.,  et~al., 2013, \mn@doi [The Astronomical Journal]
  {10.1088/0004-6256/145/6/145}, 145, 145

\bibitem[\protect\citeauthoryear{Lomb}{Lomb}{1976}]{Lomb1976}
Lomb N.~R.,  1976, \mn@doi [Astrophysics and Space Science]
  {10.1007/BF00648343}, 39, 447

\bibitem[\protect\citeauthoryear{{Longstaff}, {Casewell}, {Wynn}, {Page},
  {Williams}, {Braker}  \& {Maxted}}{{Longstaff} et~al.}{2019}]{Longstaff2019}
{Longstaff} E.~S.,  {Casewell} S.~L.,  {Wynn} G.~A.,  {Page} K.~L.,  {Williams}
  P.~K.~G.,  {Braker} I.,   {Maxted} P.~F.~L.,  2019, \mn@doi [\mnras]
  {10.1093/mnras/stz127}, \href
  {https://ui.adsabs.harvard.edu/abs/2019MNRAS.484.2566L} {484, 2566}

\bibitem[\protect\citeauthoryear{{Marsh}}{{Marsh}}{2019}]{molly}
{Marsh} T.,  2019, {molly: 1D astronomical spectra analyzer} (\mn@eprint {ascl}
  {1907.012})

\bibitem[\protect\citeauthoryear{Marsh \& Horne}{Marsh \&
  Horne}{1988}]{Marsh1988}
Marsh T.~R.,  Horne K.,  1988, \mn@doi [Monthly Notices of the Royal
  Astronomical Society] {10.1093/mnras/235.1.269}, 235, 269

\bibitem[\protect\citeauthoryear{Marsh, Horne, Schlegel, Honeycutt  \&
  Kaitchuck}{Marsh et~al.}{1990}]{Marsh1990}
Marsh T.~R.,  Horne K.,  Schlegel E.~M.,  Honeycutt R.~K.,   Kaitchuck R.~H.,
  1990, \mn@doi [The Astrophysical Journal] {10.1086/169446}, 364, 637

\bibitem[\protect\citeauthoryear{Masci et~al.,}{Masci et~al.}{2019}]{Masci2019}
Masci F.~J.,  et~al., 2019, \mn@doi [Publications of the Astronomical Society
  of the Pacific] {10.1088/1538-3873/aae8ac}, 131, 018003

\bibitem[\protect\citeauthoryear{McAllister et~al.,}{McAllister
  et~al.}{2019}]{McAllister2019}
McAllister M.,  et~al., 2019, \mn@doi [Monthly Notices of the Royal
  Astronomical Society] {10.1093/mnras/stz976}, 486, 5535

\bibitem[\protect\citeauthoryear{{Meyer} \& {Meyer-Hofmeister}}{{Meyer} \&
  {Meyer-Hofmeister}}{2015}]{Mayer2015}
{Meyer} F.,  {Meyer-Hofmeister} E.,  2015, \mn@doi [\pasj]
  {10.1093/pasj/psv023}, \href
  {https://ui.adsabs.harvard.edu/abs/2015PASJ...67...52M} {67, 52}

\bibitem[\protect\citeauthoryear{Morales-Rueda, Marsh, Steeghs, Unda-Sanzana,
  Wood  \& North}{Morales-Rueda et~al.}{2003}]{Morales-Rueda2003}
Morales-Rueda L.,  Marsh T.~R.,  Steeghs D.,  Unda-Sanzana E.,  Wood J.~H.,
  North R.~C.,  2003, \mn@doi [Astronomy and Astrophysics]
  {10.1051/0004-6361:20030552}, 405, 249

\bibitem[\protect\citeauthoryear{Morrissey et~al.,}{Morrissey
  et~al.}{2007}]{Morrissey2007}
Morrissey P.,  et~al., 2007, \mn@doi [The Astrophysical Journal Supplement
  Series] {10.1086/520512}, 173, 682

\bibitem[\protect\citeauthoryear{Motch, Haberl, Guillout, Pakull, Reinsch  \&
  Krautter}{Motch et~al.}{1996}]{Motch1996}
Motch C.,  Haberl F.,  Guillout P.,  Pakull M.,  Reinsch K.,   Krautter J.,
  1996, Astronomy and Astrophysics, 307, 459

\bibitem[\protect\citeauthoryear{{Mowlavi}}{{Mowlavi}}{1999}]{Mowlavi1999}
{Mowlavi} N.,  1999, \aap, \href
  {https://ui.adsabs.harvard.edu/abs/1999A&A...350...73M} {350, 73}

\bibitem[\protect\citeauthoryear{Nather, Robinson  \& Stover}{Nather
  et~al.}{1981}]{Nather1981}
Nather R.~E.,  Robinson E.~L.,   Stover R.~J.,  1981, \mn@doi [The
  Astrophysical Journal] {10.1086/158704}, 244, 269

\bibitem[\protect\citeauthoryear{Naylor}{Naylor}{1998}]{Naylor1998}
Naylor T.,  1998, \mn@doi [Monthly Notices of the Royal Astronomical Society]
  {10.1046/j.1365-8711.1998.01314.x}, 296, 339

\bibitem[\protect\citeauthoryear{Nelemans, Yungelson  \& {Portegies
  Zwart}}{Nelemans et~al.}{2004}]{Nelemans2004}
Nelemans G.,  Yungelson L.~R.,   {Portegies Zwart} S.~F.,  2004, Monthly
  Notices of the Royal Astronomical Society, 349, 181

\bibitem[\protect\citeauthoryear{{Osaki} \& {Meyer}}{{Osaki} \&
  {Meyer}}{2003}]{Osaki2003}
{Osaki} Y.,  {Meyer} F.,  2003, \mn@doi [\aap] {10.1051/0004-6361:20030115},
  \href {https://ui.adsabs.harvard.edu/abs/2003A&A...401..325O} {401, 325}

\bibitem[\protect\citeauthoryear{{Osaki} \& {Meyer}}{{Osaki} \&
  {Meyer}}{2004}]{Osaki2004}
{Osaki} Y.,  {Meyer} F.,  2004, \mn@doi [\aap] {10.1051/0004-6361:200400093},
  \href {https://ui.adsabs.harvard.edu/abs/2004A&A...428L..17O} {428, L17}

\bibitem[\protect\citeauthoryear{Paczy\'nski}{Paczy\'nski}{1967}]{Paczynski1967}
Paczy\'nski B.,  1967, Acta Astronomica, 17, 287

\bibitem[\protect\citeauthoryear{{Pala} et~al.,}{{Pala}
  et~al.}{2020}]{Pala2020}
{Pala} A.~F.,  et~al., 2020, \mn@doi [\mnras] {10.1093/mnras/staa764}, \href
  {https://ui.adsabs.harvard.edu/abs/2020MNRAS.494.3799P} {494, 3799}

\bibitem[\protect\citeauthoryear{Patterson et~al.,}{Patterson
  et~al.}{1998}]{Patterson1998a}
Patterson J.,  et~al., 1998, \mn@doi [Publications of the Astronomical Society
  of the Pacific] {10.1086/316252}, 110, 1290

\bibitem[\protect\citeauthoryear{{Patterson} et~al.,}{{Patterson}
  et~al.}{2002}]{Patterson2002}
{Patterson} J.,  et~al., 2002, \mn@doi [\pasp] {10.1086/341696}, \href
  {https://ui.adsabs.harvard.edu/abs/2002PASP..114..721P} {114, 721}

\bibitem[\protect\citeauthoryear{Patterson et~al.,}{Patterson
  et~al.}{2005}]{Patterson2005}
Patterson J.,  et~al., 2005, \mn@doi [The Publications of the Astronomical
  Society of the Pacific] {10.1086/447771}, 117, 1204

\bibitem[\protect\citeauthoryear{Pearson}{Pearson}{2007}]{Pearson2007}
Pearson K.~J.,  2007, \mn@doi [Monthly Notices of the Royal Astronomical
  Society] {10.1111/j.1365-2966.2007.11932.x}, 379, 183

\bibitem[\protect\citeauthoryear{Podsiadlowski, Han  \&
  Rappaport}{Podsiadlowski et~al.}{2003}]{Podsiadlowski2003}
Podsiadlowski P.,  Han Z.,   Rappaport S.,  2003, \mn@doi [Monthly Notices of
  the Royal Astronomical Society] {10.1046/j.1365-8711.2003.06380.x}, 340, 1214

\bibitem[\protect\citeauthoryear{Prieto et~al.,}{Prieto
  et~al.}{2014a}]{Prieto2014}
Prieto J.~L.,  et~al., 2014a, The Astronomer's Telegram, 6475

\bibitem[\protect\citeauthoryear{Prieto et~al.,}{Prieto
  et~al.}{2014b}]{Prieto2014b}
Prieto J.~L.,  et~al., 2014b, The Astronomer's Telegram, 6475

\bibitem[\protect\citeauthoryear{Ramsay et~al.,}{Ramsay
  et~al.}{2014}]{Ramsay2014}
Ramsay G.,  et~al., 2014, \mn@doi [Monthly Notices of the Royal Astronomical
  Society] {10.1093/mnras/stt2248}, 438, 789

\bibitem[\protect\citeauthoryear{{Ramsay} et~al.,}{{Ramsay}
  et~al.}{2018}]{Ramsay2018}
{Ramsay} G.,  et~al., 2018, \mn@doi [Astronomy and Astrophysics]
  {10.1051/0004-6361/201834261}, \href
  {https://ui.adsabs.harvard.edu/abs/2018A&A...620A.141R} {620, A141}

\bibitem[\protect\citeauthoryear{Rau et~al.,}{Rau et~al.}{2009}]{Rau2009}
Rau A.,  et~al., 2009, \mn@doi [Publications of the Astronomical Society of the
  Pacific] {10.1086/605911}, 121, 1334

\bibitem[\protect\citeauthoryear{Roelofs, Groot, Nelemans, Marsh  \&
  Steeghs}{Roelofs et~al.}{2006}]{Roelofs2006}
Roelofs G. H.~A.,  Groot P.~J.,  Nelemans G.,  Marsh T.~R.,   Steeghs D.,
  2006, \mn@doi [Monthly Notices of the Royal Astronomical Society]
  {10.1111/j.1365-2966.2006.10718.x}, 371, 1231

\bibitem[\protect\citeauthoryear{Roelofs, Rau, Marsh, Steeghs, Groot  \&
  Nelemans}{Roelofs et~al.}{2010}]{Roelofs2010}
Roelofs G. H.~A.,  Rau A.,  Marsh T.~R.,  Steeghs D.,  Groot P.~J.,   Nelemans
  G.,  2010, \mn@doi [The Astrophysical Journal]
  {10.1088/2041-8205/711/2/L138}, 711, L138

\bibitem[\protect\citeauthoryear{Ruiz, Rojo, Garay  \& Maza}{Ruiz
  et~al.}{2001}]{Ruiz2001}
Ruiz M.~T.,  Rojo P.~M.,  Garay G.,   Maza J.,  2001, \mn@doi [The
  Astrophysical Journal] {10.1086/320578}, 552, 679

\bibitem[\protect\citeauthoryear{Rykoff et~al.,}{Rykoff
  et~al.}{2004}]{Rykoff2004}
Rykoff E.~S.,  et~al., 2004, Information Bulletin on Variable Stars, 5559

\bibitem[\protect\citeauthoryear{Savonije, de Kool  \& van~den Heuvel}{Savonije
  et~al.}{1986}]{Savonije1986}
Savonije G.~J.,  de Kool M.,   van~den Heuvel E. P.~J.,  1986, Astronomy and
  Astrophysics, 155, 51

\bibitem[\protect\citeauthoryear{Scargle}{Scargle}{1982}]{Scargle1982}
Scargle J.~D.,  1982, \mn@doi [The Astrophysical Journal] {10.1086/160554},
  263, 835

\bibitem[\protect\citeauthoryear{Shappee et~al.,}{Shappee
  et~al.}{2014}]{Shappee2014}
Shappee B.~J.,  et~al., 2014, \mn@doi [The Astrophysical Journal]
  {10.1088/0004-637X/788/1/48}, 788, 48

\bibitem[\protect\citeauthoryear{Smak}{Smak}{1981}]{Smak1981}
Smak J.,  1981, Acta Astronomica, 31, 395

\bibitem[\protect\citeauthoryear{Solheim}{Solheim}{2010}]{SolheimAMCVn}
Solheim J.-E.,  2010, \mn@doi [Publications of the Astronomical Society of the
  Pacific] {10.1086/656680}, 122, 1133

\bibitem[\protect\citeauthoryear{Steeghs, Marsh, Barros, Nelemans, Groot,
  Roelofs, Ramsay  \& Cropper}{Steeghs et~al.}{2006}]{Steeghs2006}
Steeghs D.,  Marsh T.~R.,  Barros S. C.~C.,  Nelemans G.,  Groot P.~J.,
  Roelofs G. H.~A.,  Ramsay G.,   Cropper M.,  2006, \mn@doi [The Astrophysical
  Journal] {10.1086/506343}, 649, 382

\bibitem[\protect\citeauthoryear{Szkody et~al.,}{Szkody
  et~al.}{2005}]{Szkody2005}
Szkody P.,  et~al., 2005, \mn@doi [The Astronomical Journal] {10.1086/429595},
  129, 2386

\bibitem[\protect\citeauthoryear{Thorstensen, Fenton, Patterson, Kemp, Krajci
  \& Baraffe}{Thorstensen et~al.}{2002}]{Thorstensen2002}
Thorstensen J.~R.,  Fenton W.~H.,  Patterson J.~O.,  Kemp J.,  Krajci T.,
  Baraffe I.,  2002, \mn@doi [The Astrophysical Journal] {10.1086/339905}, 567,
  L49

\bibitem[\protect\citeauthoryear{Uthas et~al.,}{Uthas et~al.}{2011}]{Uthas2011}
Uthas H.,  et~al., 2011, \mn@doi [Monthly Notices of the Royal Astronomy
  Society] {10.1111/j.1745-3933.2011.01061.x}, 414, L85

\bibitem[\protect\citeauthoryear{VanderPlas}{VanderPlas}{2018}]{VanderPlas2018}
VanderPlas J.~T.,  2018, \mn@doi [The Astrophysical Journal Supplement Series]
  {10.3847/1538-4365/aab766}, 236, 16

\bibitem[\protect\citeauthoryear{Vernet et~al.,}{Vernet
  et~al.}{2011}]{Vernet2011}
Vernet J.,  et~al., 2011, \mn@doi [Astronomy and Astrophysics]
  {10.1051/0004-6361/201117752}, 536, A105

\bibitem[\protect\citeauthoryear{Wang, Steeghs, Casares, Charles,
  Mu{\~{n}}oz-Darias, Marsh, Hynes  \& O'Brien}{Wang et~al.}{2017}]{Wang2017}
Wang L.,  Steeghs D.,  Casares J.,  Charles P.~A.,  Mu{\~{n}}oz-Darias T.,
  Marsh T.~R.,  Hynes R.~I.,   O'Brien K.,  2017, \mn@doi [Monthly Notices of
  the Royal Astronomical Society] {10.1093/mnras/stw3312}, 466, 2261

\bibitem[\protect\citeauthoryear{Wang, Steeghs, Galloway, Marsh  \&
  Casares}{Wang et~al.}{2018}]{Wang2018}
Wang L.,  Steeghs D.,  Galloway D.~K.,  Marsh T.,   Casares J.,  2018, \mn@doi
  [Monthly Notices of the Royal Astronomical Society] {10.1093/mnras/sty1441},
  478, 5174

\bibitem[\protect\citeauthoryear{Warner}{Warner}{1995}]{Warner1995}
Warner B.,  1995, {Cataclysmic Variable Stars}.
Cambridge University Press, Cambridge

\bibitem[\protect\citeauthoryear{Wood, Horne, Berriman, Wade, O'Donoghue  \&
  Warner}{Wood et~al.}{1986}]{Wood1986a}
Wood J.,  Horne K.,  Berriman G.,  Wade R.,  O'Donoghue D.,   Warner B.,  1986,
  \mn@doi [Monthly Notices of the Royal Astronomical Society]
  {10.1093/mnras/219.3.629}, 219, 629

\bibitem[\protect\citeauthoryear{Woudt \& Warner}{Woudt \&
  Warner}{2011}]{Woudt2011}
Woudt P.~A.,  Warner B.,  2011, The Astronomer's Telegram, 3705

\bibitem[\protect\citeauthoryear{Woudt, Warner, de Bude, Macfarlane, Schurch
  \& Zietsman}{Woudt et~al.}{2012}]{Woudt2012}
Woudt P.~A.,  Warner B.,  de Bude D.,  Macfarlane S.,  Schurch M. P.~E.,
  Zietsman E.,  2012, \mn@doi [Monthly Notices of the Royal Astronomical
  Society] {10.1111/j.1365-2966.2012.20476.x}, 421, 2414

\bibitem[\protect\citeauthoryear{Wright et~al.,}{Wright
  et~al.}{2010}]{Wright2010}
Wright E.~L.,  et~al., 2010, \mn@doi [The Astronomical Journal]
  {10.1088/0004-6256/140/6/1868}, 140, 1868

\bibitem[\protect\citeauthoryear{Yungelson}{Yungelson}{2008}]{Yungelson2008}
Yungelson L.~R.,  2008, \mn@doi [Astronomy Letters]
  {10.1134/S1063773708090053}, 34, 620

\makeatother
\end{thebibliography}






\bsp	
\label{lastpage}
\end{document}